%% file: Draft.tex
\documentclass[journal,draftcls,onecolumn,12pt,twoside]{IEEEtranTCOM}

\usepackage{graphicx}
\usepackage{epstopdf}
\usepackage{amsmath, amssymb, amsthm, amsfonts}
\usepackage[font=footnotesize]{subfig}
\usepackage{mathtools}
\usepackage{tikz}
\usetikzlibrary{shapes}
\usepackage{pgfplots}
\tikzset{>=latex}

\usepackage{cite}
\usepackage{multicol}

\usepackage[ruled,vlined,linesnumbered]{algorithm2e}
\usepackage{algpseudocode}

\usepackage{bbm}

\usepackage{mathrsfs}

\usetikzlibrary{patterns}
\usetikzlibrary{automata, positioning}

\renewcommand\IEEEkeywordsname{Index terms}

\newtheorem{lemma}{Lemma}
\newtheorem{theorem}{Theorem}
\newtheorem{proposition}{Proposition}

\newtheorem*{remark}{Remark}

\begin{document}
	
\bstctlcite{IEEEexample:BSTcontrol}

\title{The Age of Information in a Discrete Time Queue: Stationary Distribution and Non-linear Age Mean Analysis}

\author{Antzela~Kosta,~\IEEEmembership{Member,~IEEE,}
	Nikolaos~Pappas,~\IEEEmembership{Member,~IEEE,}
	Anthony~Ephremides,~\IEEEmembership{Life Fellow,~IEEE,}
	and~Vangelis~Angelakis,~\IEEEmembership{Senior Member,~IEEE}% <-this % stops a space
	\thanks{This work extends the preliminary study in \cite{Kosta20_ICC}. 
		
	A. Kosta, N. Pappas, and V. Angelakis are with the Department of Science and Technology, Link{\"o}ping University, Norrk{\"o}ping SE-60174, Sweden (email: antzela.kosta@liu.se, nikolaos.pappas@liu.se, vangelis.angelakis@liu.se). 
	A. Ephremides is with the Department of Electrical and Computer Engineering and the Institute for System Research, University of Maryland, College Park, MD 20740, USA (email: etony@ece.umd.edu), also collaborating with the Department of Science and Technology, Link{\"o}ping University, Norrk{\"o}ping SE-60174, Sweden.	
		
	The research leading to these results has been partially funded by the
	European Union's Horizon 2020 research and innovation programme under the Marie Sklodowska-Curie Grant Agreement No. 642743 (WiVi-2020). 
	In addition, this work was supported in part by ELLIIT and the Center for Industrial Information Technology (CENIIT).
	The work of Anthony Ephremides is supported by the U.S. Office of Naval Research under Grant ONR 5-280542, the U.S. National Science Foundation under Grants CIF 5-243150, Nets 5-245770, and CIF 5-231912, and the Swedish Research Council (VR).}}% 

\maketitle

\vspace{-10mm}
\begin{abstract}
In this work, we investigate information freshness in a status update communication system consisting of a source-destination link.	
Initially, we study the properties of a sample path of the age of information (AoI) process at the destination. 
We obtain a general formula of the stationary distribution of the AoI, under the assumption of ergodicity.
We relate this result to a discrete time queueing system and provide a general expression of the generating function of AoI in relation with the system time and the peak age of information (PAoI) metric.
Furthermore, we consider three different single-server system models and we obtain closed-form expressions of the generating functions and the stationary distributions of the AoI and the PAoI.
The first model is a first-come-first-served (FCFS) queue, the second model is a preemptive last-come-first-served (LCFS) queue, and the last model is a bufferless system with packet dropping.
We build upon these results to provide a methodology for analyzing general non-linear age functions for this type of systems, using representations of functions as power series.
\end{abstract}
%\vspace*{-0.5cm}

\IEEEpeerreviewmaketitle
\vspace*{-0.4cm}
\section{Introduction}
In communication systems, it is common to deal with time critical information that needs to be transmitted from the generation point to a remote destination in the network.
To address timeliness requirements, the notion of age of information (AoI) has been introduced to quantify the freshness of the received information at the destination \cite{NET-060,Abd19_Magazine,Sun19_Book}.
At any moment, the AoI at the destination is the time that elapsed since the last received status update was generated by the source.

Systems with different availabilities of resources have been modeled through different queueing models and the time average AoI was derived.
In \cite{Kaul12_INFOCOM}, the M/M/1, the M/D/1, and the D/M/1 queues were studied under the first-come-first-served (FCFS) discipline. 
The last-come-first-served (LCFS) queue discipline with or without the ability to preempt the packet in service has been considered in \cite{Kaul12_CISS,Najm16_ISIT,Bedewy16_ISIT,Yates18_INFOCOM,Najm18_INFOCOM}.
The effect of the buffer size and the available number of servers has been studied in \cite{Costa16,Soysal19_arXiv,Kosta19_ISIT,Kosta19_JCN,Kam16,Bedewy19_transactionsIT}.
The problem of minimizing AoI over the space of all inter-generation and service time distributions, in various continuous time and discrete time queueing systems, such as the FCFS G/G/1, the LCFS G/G/1, and the G/G/$\infty$, has been investigated in \cite{Tripathi19_arXiv,Talak18_Determinacy}.
In \cite{Chen16_ISIT}, the authors consider the peak age of information (PAoI) in an M/M/1 queueing system with packet delivery errors.

Furthermore, a number of works depart from the average AoI and consider a complete characterization of the AoI distribution.
The probability distribution of AoI is an important metric that highlights the AoI key characteristics for a given system and provides performance guarantees.
In \cite{Yates18_arXivSHS} new tools such as stochastic hybrid systems (SHS) are developed to analyze AoI moments and the moment generating function (MGF) of an AoI process in networks.
Closely related to this work is \cite{Inoue19_transactionsIT} that provides 
a general formula of the stationary distribution of AoI for a wide class of continuous-time single server queues with different disciplines.
The distribution of AoI for the GI/GI/1/1 and GI/GI/1/2* systems, under non-preemptive scheduling is considered in \cite{Champati19_INFOCOM}. 
In \cite{Kesidis19_arXiv} the authors characterize the AoI distribution in bufferless systems.
Delay and PAoI violation guarantees are studied in 
\cite{Devassy19_JSAC} for the reliable transmission of short packets over a wireless channel.

The AoI is determined by two factors \emph{(i)} the processing/transmission delay and \emph{(ii)} the process that generates status updates.
To capture both the information characteristics of the source and these factors, it is meaningful to modify the definition of AoI to a non-linear cost function.
The aim is to penalize the absence of updates at the destination according to the source characteristics by a non-negative, monotonically increasing function.

The works in \cite{Kosta17_ISIT,Kosta19_arXiv} aim to expand the concept of information ageing by introducing the cost of update delay (CoUD) metric to provide a flexible measure of having stale information at the destination depending on the autocorrelation
properties of the source.
In \cite{Sun2017_transactions,Sun18_JCN}, sampling for data freshness is considered, and so-called age penalty and utility functions are employed to describe the level of dissatisfaction for having aged status updates at the destination.
In \cite{Sun18_SPAWC}, the authors use the mutual information between the real-time source value and the delivered samples at the receiver to quantify the freshness of the information contained in the delivered samples.
In \cite{Sun19_transactions} it was proven that sampling a Wiener process to minimize AoI is not optimal for minimizing the estimation error.
Recently, a relationship between a non-linear
function of the AoI and the estimation error of the Ornstein-Uhlenbeck (OU) process was investigated in \cite{Ornee19_WiOpt}.
To the best of our knowledge, the probability distribution of
AoI and the non-linear age mean in discrete time queueing systems are not yet covered in the existing literature.

\subsection{Contributions}

In this work, we investigate information freshness in a source-destination communication link.
Following the approach of \cite{Inoue19_transactionsIT} that studies the distribution of the AoI process, 
we derive a general formula for the stationary distribution
of the AoI, that holds for a wide class of discrete time status update systems.
Studying the properties of a sample path of the AoI we can have a better and deeper understanding of the underlying AoI properties under general assumptions.
More specifically, we determine the relation among the AoI, the system delay, and the PAoI.
We invoke ergodicity that has been extensively used in the AoI literature and we apply our results to a discrete time queueing system.
To that extent, we obtain general formulas of the z-transforms of the AoI and PAoI metrics.
To illustrate the applicability of the results we consider an FCFS queue, a preemptive LCFS queue, and a bufferless system with packet dropping, and we derive the probability distribution function (PDF) and the probability mass function (pmf) of AoI and PAoI, that can be utilized to provide performance guarantees.
Moreover, we provide a methodology for analyzing general non-linear age functions for this type of systems.
We illustrate how our results can be used to obtain closed-form expressions of the time average performance of non-linear functions of the AoI, providing some examples. 
Doing so, we use representations of functions as power series and develop an algorithm that approached the exact time-average performance up to the desired accuracy.
Furthermore, we consider the problem of maximizing freshness by obtaining the optimal arrival probability at the queue.
Our results are general enough to be utilized to a variety of settings in terms of the queueing discipline, the arrival and service process, and the non-linear cost function at the destination.

\section{The AoI Sample Path in a General Setting}

We consider a stochastic system consisting of a point-to-point communication link with a single transmitter node (source) that sends status updates to a single receiver node (destination).
Time is assumed to be slotted.
The AoI of the transmitter node at the receiver node is defined as the random process
\begin{equation}
	\Delta_t = t-u(t), \quad t \in \mathbb{Z}^*,
	\label{eq:age}
\end{equation}
where $u(t)$ is the time-stamp of the most recently received status update.

Any sample path of the AoI process $\Delta_t$ can be constructed as follows.
Let $\{t'_n, n \geq 0\}$ be a deterministic point process, with $t'_0=0$ and $t'_n < t'_{n+1}<\infty$.
We interpret $t'_n$ as the times at which the status updates are received at the destination.
At time $t$, the number of events in $[0,t]$ is denoted by
$N(t)=\max\{n: t'_n \leq t\}$, $t \geq 0$. 
We assume that $t'_n \to \infty$ as $n \to \infty$, so that there is a finite number of receptions in any finite time interval, and we note that, since $t'_n <\infty$ for all $n\geq 0$, we have that $N(t) \to \infty$ as $n \to \infty$.
Associated with each point $t'_n$ is the mark $T_n = \{\Delta_{t'_n}, n \geq 0\}$ denoting the value of AoI immediately after receiving the $n$th status update.
Then, $\{(t'_n,T_n), n\geq0\}$ denotes the marked point process of AoI on $[0,\infty)\times[0,\infty)$.
The AoI process is non-negative, piece-wise non-decreasing, right-continuous, with discontinuous jumps at times $t'_n$.
A sample path of the AoI process is shown in Fig.~\ref{fig:age_vs_time_slotted}.

The AoI process $\Delta_t$ is thus determined completely by $\{(t'_n,T_n), n\geq0\}$ as follows
\begin{equation}
	\Delta_t= T_{n-1} + (t-t'_{n-1}) , \quad t \in[t'_{n-1},t'_n), \quad n\geq1.
	\label{eq:age_def}
\end{equation}
Moreover, we define the PAoI as the value of AoI achieved immediately before receiving the $n$th update 
\begin{equation}
	A_n=T_{n-1}+(t'_{n}-t'_{n-1}), \quad n\geq1.
	\label{eq:peak_age_def}
\end{equation}

Some notations are given in TABLE \ref{tab:notation} for convenience.

\begin{table}[t]
	\centering
	\begin{tabular}{|c|c||c|c|}
		\hline
		Symbol & Description & Symbol & Description \\
		\hline
		$\Delta_t$ & the AoI random process & $A_n$ & the PAoI random process \\
		\hline
		$T_n$ & the system time of the $n$th update & $Y_n$ & the interarrival time of the $n$th update\\
		\hline
		$t_n$ & the sampling time of the $n$th update & $t'_n$ & the reception time of the $n$th update\\
		\hline
		$\lambda^{\dagger}$ & the arrival rate of $\Delta_t$ & $\mu$ & the service rate of the queue server \\
		\hline
		$\Delta^{\dagger}(x)$, $A^{\dagger}(x)$, $T^{\dagger}(x)$ & asymptotic frequency distributions & $S_n$ & the service time of the $n$th update \\
		\hline
		$\Delta^{*}(z)$, $A^{*}(z)$, $T^{*}(z)$, $S^{*}(z)$  & z-transforms & $z$ & the z-transform complex variable \\
		\hline
		$\Delta(x)$, $A(x)$, $T(x)$ & probability distribution functions & $\lambda$ & the queue average arrival rate\\
		\hline
		$P_{\Delta}(x)$, $P_A(x)$, $P_T(x)$ & probability mass functions & $C_t$ & a function of the AoI random process \\
		\hline
		$C$, $C_{peak}$ & the time average CoUD and PCoUD & $\alpha$ & a non-negative parameter \\
		\hline
		$f_s(t)$ & a class of non-decreasing functions & $\Phi$ & the Lerch transcendent \\
		\hline
		$\theta$, $p$ & probabilities & $\lambda_e$ & the effective arrival rate \\
		\hline
	\end{tabular}
	\caption{Notation}
	\label{tab:notation}
\end{table}

\subsection{Main Result}

Consider a fixed sample path of $\{(t'_n,T_n), n\geq0\}$ where the quantities are deterministic.
Let $\Delta^{\dagger}(x)$, $A^{\dagger}(x)$, and $T^{\dagger}(x)$, denote the asymptotic frequency distributions \cite{ElTaha1999sample} of $\{\Delta_t, t\geq 0\}$, $\{A_n, n\geq1\}$, and $\{T_n, n\geq1\}$, respectively. When the limits exist these are given by 

\begin{equation}
	\Delta^{\dagger}(x)= \lim_{\mathcal{T} \to \infty}   \frac{1}{\mathcal{T}}\sum_{t=0}^{\mathcal{T}} \mathbbm{1}_{\{ \Delta_t\leq x\}}, \quad x \geq 0,
	\label{eq:age_fre_distribution}
\end{equation}

\begin{equation}
	A^{\dagger}(x)= \lim_{N \to \infty}  \frac{1}{N}\sum_{n=1}^{N} \mathbbm{1}_{\{ A_n\leq x\}}, \quad x \geq 0,
	\label{eq:peakage_fre_distribution}
\end{equation}

\begin{equation}
	T^{\dagger}(x)= \lim_{N \to \infty} \frac{1}{N}\sum_{n=1}^{N} \mathbbm{1}_{\{ T_n\leq x\}}, \quad x \geq 0,
	\label{eq:T_fre_distribution}
\end{equation}
where $\mathbbm{1}_{\{\cdot\}}$ is the indicator function.

The following lemma is a sample-path analogue of the elementary renewal theorem.
\begin{lemma}[\hspace{1sp}\cite{ElTaha1999sample}, Lemma 1.1]\label{lemma1}
	Let $0<\lambda^{\dagger}<\infty$. 
	%Then $t^{-1} N(t) \to \lambda^{\dagger}$ as $t \to \infty$ if and only if $n^{-1} t'_n \to 1/\lambda^{\dagger}$ as $n \to \infty$.  	
	Then, $\lim_{t \to \infty} \frac{N(t)}{t} = \lambda^{\dagger} \quad$  if and only if $ \quad \lim_{n \to \infty} \frac{t'_n}{n} = \frac{1}{\lambda^{\dagger}}$.  
\end{lemma}
Next, we have the following theorem.
\begin{theorem}\label{theorem1}
	If the limits \eqref{eq:peakage_fre_distribution} and \eqref{eq:T_fre_distribution} exist for each $x\geq0$, then the limit in \eqref{eq:age_fre_distribution} also exists and it is given by 
	\begin{equation}
		\Delta^{\dagger}(x)=\lambda^{\dagger}  \sum_{u=0}^{x}  (T^{\dagger}(u) - A^{\dagger}(u)).
		\label{eq:main_result}
	\end{equation}
	\begin{proof} The proof is given in Appendix~\ref{Appendix_A'}. \end{proof}	
\end{theorem}

%\afterpage{%
\begin{figure}[t!]
	\centering
	\input{age_vs_time_slotted.tex}
	\caption{A sample path of the AoI process.}
	\label{fig:age_vs_time_slotted}
\end{figure}%}

\vspace*{-0.4cm}
\section{The case of a Discrete Time Queue}

In this section, we relate the sample-path of the previous section to a stationary, ergodic queueing system.
We consider that the transmitter node has a buffer of infinite capacity to store incoming status updates in the form of packets.
These packets are then sent through an error-free channel to the destination. 
Packets have equal length and time is divided into slots such that the transmission time of a packet from the buffer to the destination is equal to one slot.
Note that the queue follows an early departure-late arrival model, that is, 
departures take place at the beginning of a slot and arrivals at the end of the slot.
Each such packet is said to provide a \emph{status update} and these two terms are used interchangeably.
The status update interarrival times are modeled as independent and identically distributed (i.i.d.) random variables, with the average arrival rate denoted by $\lambda$.
Moreover, we consider a general service process where the service times are i.i.d. with average service rate $\mu$, and a single server.
We specify the interarrival distribution and service
distribution in the next sections.

Consider that the $n$th status update is generated at time $t_{n}$, delivered through the transmission system, and received by the destination at time $t_{n}^{'}$.
Then, we denote by $T_{n} = t'_{n}  - t_{n}$ the system time of update $n$.
This corresponds to the sum of the queueing time and the queue service time.
The interarrival time of update $n$ is defined as the
random variable $Y_{n} = t_{n} - t_{n-1}$.
Then, the PAoI achieved immediately before receiving the $n$th update in \eqref{eq:peak_age_def} can be defined alternatively as
\begin{equation}
	A_n=Y_n+T_n.
\label{eq:PAoI_def}
\end{equation}

Let the PDF of the random variables $T$ and $A$ be denoted by $T(x)$ and $A(x)$, respectively.
The stationary distribution $\Delta(x)$, of the AoI is defined as the long-run fraction of time in which the AoI is less than or equal to an arbitrary fixed value $x$, i.e., 
\begin{equation}
	\Delta(x)= \lim_{\mathcal{T} \to \infty}   \frac{1}{\mathcal{T}}\sum_{t=0}^{\mathcal{T}} \mathbbm{1}_{\{ \Delta_t\leq x\}}.
	\label{eq:age_distribution}
\end{equation}
Furthermore, let $\Delta^*(z)$, $A^*(z)$, and $T^*(z)$, denote the z-transforms (also known as the generating functions) of the AoI, PAoI, and system time, respectively.
The z-transform for a function $\gamma(n)$ is defined as 
\begin{equation}
	\Gamma^*(z) = \sum_{n=0}^{\infty} \gamma(n) z^n.
	\label{eq:z_transform}
\end{equation}

\begin{lemma}\label{lemma2}
	For stationary and ergodic systems $\Delta^{\dagger}(x)=\Delta(x)$,  $A^{\dagger}(x)=A(x)$, and $T^{\dagger}(x)=T(x)$, with probability 1.
	\begin{proof} 
		%The equations follow from the ergodic theorem in \cite[Theorem A.4]{ElTaha1999sample}.
		We invoke the ergodic theorem in \cite[Theorem A.4]{ElTaha1999sample} to guarantee that the asymptotic frequency distributions exist and coincide with the corresponding stationary probabilities.		
	\end{proof}	
\end{lemma}

Next, we have the following theorem.
\begin{theorem}\label{theorem2}
	The z-transform of the AoI of the source at destination is given by  
	\begin{equation}
		\Delta^{\ast}(z)= \lambda \frac{T^*(z)-A^*(z)}{(1-z)/z}.
		\label{eq:AoI_z_transform}
	\end{equation}
	%\begin{proof} The proof is straightforward from Theorem 1 and properties of the z-transform. \end{proof}	
	\begin{proof} The proof is given in Appendix~\ref{Appendix_B'}. \end{proof}	
\end{theorem}

%FIFO

\section{The  FCFS  Queue Discipline}
This section considers an FCFS queue over the space of all interarrival and service time distributions.
We proceed by deriving a general formula of the z-transform of the PAoI $A^{\ast}(z)$.
We observe that the random variables $Y_n$ and $T_n$ in \eqref{eq:PAoI_def} are dependent and provide an alternative definition of PAoI that is
\begin{equation}
	A_n= \max(Y_n,T_{n-1}) +S_n,
	\label{eq:PAoI_def_alt}
\end{equation}
where $S_n$ denotes the service time of the $n$th update.
Then, the random variables $Y_n$ and $T_{n-1}$ are independent and we define $Z=\max(Y_n,T_{n-1})$.
The probability distribution of $Z$ is given by 
\begin{align}
	Z(x) &= \text{Pr} (Z \leq x) = \text{Pr} (\max(Y_n,T_{n-1}) \leq x) = \notag \\ 
	%&= \text{Pr} (Y_n \leq x \quad \text{and} \quad T_{n-1} \leq x) = \notag \\
	&=\text{Pr} (Y_n \leq x , T_{n-1} \leq x)=\text{Pr} (Y_n \leq x) \text{Pr} (T_{n-1} \leq x) = Y(x) T(x),
	\label{eq:Z_CDF}
\end{align}
$x >0$. Moreover, the pmf of $Z$ is given by
\begin{align}
	P_Z(x) = \text{Pr} (Z=x)= Z(x) - Z(x-1)  
	= Y(x) T(x) - Y(x-1) T(x-1).
	\label{eq:Z_pmf}
\end{align}
As a result, the z-transform of the PAoI is obtained as 
\begin{align}
	A^{\ast}(z) = Z^{\ast}(z) S^{\ast}(z)   
	= \left[ \sum_{n=1}^{\infty} \big( Y(n) T(n) - Y(n-1) T(n-1) \big) z^n \right] S^{\ast}(z).
	\label{eq:PAoI_z_transform}
\end{align}

\section{The FCFS Geo/Geo/1 queue}
Consider a discrete time Geo/Geo/1 queue, where the arrival process is modeled as Bernoulli with average probability $\lambda \in \left( 0,1 \right)$.
The probability distribution of time until successful delivery is assumed to be geometric with mean $\mathbb{E}[S]=1/\mu$ slots, where $\mu$ is referred as the service probability.
In the following theorem we obtain the generating functions of PAoI and AoI.

\begin{theorem}\label{theorem3}
	The z-transforms of the PAoI and the AoI for the Geo/Geo/1 queue with an FCFS queue discipline are given by
	\begin{equation}
		A^{\ast}(z)	=\frac{\lambda \mu (\mu-\lambda) z^2 (1-(1-\mu) z^2)}{(1-(1-\lambda) z) (1-(1-\mu) z)^2 (1-\lambda-(1-\mu)z)},
		\label{eq:PAoI_gen_func}
	\end{equation}
	and
	\begin{equation}
		\Delta^{\ast}(z)= \frac{\lambda (\mu-\lambda) z^2 (1-(1-\mu) z (2-\lambda- (1-\lambda-\mu)z))}{(1-(1-\lambda) z) (1-(1-\mu) z)^2 (1-\lambda-(1-\mu)z)},
		\label{eq:AoI_gen_func}
	\end{equation}	
	respectively.
	\begin{proof} The proof is given in Appendix~\ref{Appendix_C'}. \end{proof}	
\end{theorem}

As a result, taking the inverse z-transform we can obtain the pmfs of the PAoI and the AoI as
\begin{align}
	P_A(x) &= \text{Pr} (A=x)= \mu \Bigg(
	\frac{(\mu-\lambda) \rho^{x-1}}{\lambda (1-\mu)}
	+\mu (1-x) (1-\mu)^{x-2} +
	\notag \\ 
	& 
	+\frac{\lambda (1-\lambda)^{x-1}}{\mu-\lambda} 
	+\frac{\left(\lambda^2 (\mu-2)+2 \lambda \mu-\mu^2\right) (1-\mu)^{x-2}}{\lambda (\mu-\lambda)} 
	\Bigg),
	\label{eq:A_pmf}
\end{align}
and
\begin{align}
	P_{\Delta}(x) &= \text{Pr} (\Delta=x)=
	\frac{(\mu-\lambda) \rho^{x-1}}{1-\mu}
	+\lambda \mu (1-x) (1-\mu)^{x-2} +\notag \\ 
	& +\frac{\lambda \mu (1-\lambda)^{x-1}}{\mu-\lambda} + \frac{\left(\lambda^2-\lambda \mu (\mu+1)+\mu^2\right) (1-\mu)^{x-2}}{\lambda-\mu},
	\label{eq:Delta_pmf}
\end{align}
where we define the ratio $\rho=\frac{1-\mu}{1-\lambda}$.

Furthermore, we derive the PDFs of the PAoI and the AoI as follows
\begin{align}
	&A(x) = \text{Pr} (A \leq x) =  
	\frac{\mu (\lambda-1) \rho^{x} }{\lambda (1-\mu)} + \frac{1}{\lambda (1-\mu) (\mu-\lambda)}
	\notag \\ 
	& \Big((1-\mu)^x \big(\lambda^2 (1-\mu x)+\lambda \mu (\mu (x-1)-1)+\mu^2\big) +\lambda (\mu-1) \left(\mu \left((1-\lambda)^x-1\right)+\lambda\right)\Big),
	\label{eq:A_PDF}
\end{align}
and
\begin{align}
	\Delta(x) &= \text{Pr} (\Delta \leq x) =
	\frac{1}{(1-\mu) (\mu-\lambda)} \rho^{x} \Bigg( -\mu + \lambda^2 \left((1-x) (1-\mu)^x \rho^{-x}-1\right) \notag \\ 
	& - \mu \left(\mu-(1-\mu)^x-1\right) \rho^{-x}+(\mu-1) \mu (1-\lambda)^x \rho^{-x} + \notag \\
	& \lambda \left(\left(\mu (x-2) (1-\mu)^x+\mu-1\right) \rho^{-x} \hspace{-0.7mm}+\mu+1\right) \hspace{-0.7mm} \Bigg).
	\label{eq:Delta_PDF}
\end{align}
Note that in this context of discrete time queueing systems $x$ cannot be less that one (time-slot).
These can be utilized for quantifying the AoI or PAoI violation probability that can in turn provide AoI and PAoI performance guarantees for the system.

\section{The  preemptive LCFS  Queue Discipline}
In this section, we consider an LCFS queue with preemptive service, where newly generated updates are prioritized by interrupting the packet in service. 
Every packet enters the service immediately after its generation and it might either complete service or be preempted by a new arrival and wait.
We refer to a packet which receives service and carries the newest information compared to the packets arriving at the
destination prior to it as an \emph{informative packet}.
A non-informative packet is one that is
rendered obsolete. 
Note that the treatment of obsolete packets (i.e., discarding them or not) does not affect the AoI performance in this queue discipline.

We define $\bar{S}(x) = 1-S(x)$ and $\bar{Y}(x) = 1-Y(x)$, $x>0$, to be the  complementary cumulative distribution functions (CCDFs) of $S$ and $Y$, respectively.
Moreover, let 
\begin{equation}
\theta = \text{Pr} (Y \leq S) =\sum_{n=1}^{\infty} Y(n) P_S(n) =\sum_{n=1}^{\infty} \bar{S}(n) P_Y(n), 
\label{eq:zeta_def}
\end{equation}	
be the probability that an arriving packet becomes obsolete.
Note that we consider the interarrival and service times to be independent.
Note that $\theta=1$ means that packets are preempted in service with probability 1, and $\theta=0$ corresponds to all packets being informative.
In the latter case, the PAoI in \eqref{eq:PAoI_def} is simply given by the  convolution of two independent random variables that represent the interarrival time of an update and the service time of the same update.
In the z-domain this corresponds to $A^{\ast}(z)=Y^{\ast}(z) S^{\ast}(z)$.

Next, we define the conditional random variable $S_{<Y}$ of a service time given that it is smaller than an interarrival time, the conditional random variable $Y_{<S}$ of an interarrival time given that it is smaller than a service time, and the conditional random variable $Y_{>S}$ of an interarrival time given that it is greater than a service time.
Then, the corresponding z-transforms are defined as follows
%Next, we define the following z-transforms of conditional random variables
\begin{equation}
S^{\ast}_{<Y}(z) = \frac{1}{1-\theta} \sum_{n=1}^{\infty} \bar{Y}(n) P_S(n) z^n,
\label{eq:SlessthanY_def}
\end{equation}	
\begin{equation}
Y^{\ast}_{<S}(z) = \frac{1}{\theta} \sum_{n=1}^{\infty} \bar{S}(n) P_Y(n) z^n, 
\label{eq:YlessthanS_def}
\end{equation}	
\begin{equation}
Y^{\ast}_{>S}(z) = \frac{1}{1-\theta} \sum_{n=1}^{\infty} S(n) P_Y(n) z^n. 
\label{eq:YmorethanS_def}
\end{equation}	

We proceed by deriving a general formula of the z-transform of the PAoI $A^{\ast}(z)$.
We observe that the value of AoI achieved immediately before receiving the $n$th informative update can be described as a summation of three independent parts.
The first part represents the first packet arrival after the successful delivery of the $n-1$th status update. 
Then, we have $m$ ($m=1,2, \dots$) interarrival times (of the $i$th non-informative update) given that each one of them is smaller than the service time, and finally the service time of the $n$th informative status update. 
As a result, we have 
\begin{equation}
A_n = Y_{>S} + \sum_{i=1}^{m-1} Y^{[i]}_{<S} + S_{<Y}.
\label{eq:PAoI_preemption}
\end{equation}	
The $m$th generated packet becomes informative with probability $p_m= (1-\theta) \theta^{m-1}$. 
This implies that 
\begin{align}
A^{\ast}(z)  = Y^{\ast}_{>S}(z) \sum_{m=1}^{\infty} p_m [Y^{\ast}_{<S}(z)]^{m-1}  S^{\ast}_{<Y}(z) = Y^{\ast}_{>S}(z)  \frac{1-\theta}{1-\theta Y^{\ast}_{<S}(z)}  S^{\ast}_{<Y}(z).
\label{eq:PAoI_z_preemption}
\end{align}	

To derive a general formula of the z-transform of the AoI $\Delta^{\ast}(z)$, we refer to the time average rate of informative packets as the effective rate and define it as
\begin{equation}
\lambda_e = \left( 1-\theta \left(1-\frac{1}{\mathbb{E}[S]} \right)\right) \frac{1}{\mathbb{E}[Y]}.
\label{eq:effective_arrival}
\end{equation}	
Moreover, we note that the system time of a packet is given by $T^{\ast}(z)=S^{\ast}_{<Y}(z)$.
Finally, we utilize Theorem~\ref{theorem2} to obtain 
\begin{equation}
\Delta^{\ast}(z)= \lambda_e \frac{S^{\ast}_{<Y}(z)-A^*(z)}{(1-z)/z}.
\label{eq:th2_preemption}
\end{equation}	

Note that informative packets are transmitted from the source to the destination in a first-in-first-out (FIFO) manner.
As a result, Theorem~\ref{theorem2} is applicable to preemptive LCFS queues.

\section{The preemptive LCFS  Geo/Geo/1 Queue}
Consider a discrete time LCFS Geo/Geo/1 queue with preemptive service, where the arrival process is modeled as Bernoulli with average probability $\lambda \in \left( 0,1 \right)$.
The probability distribution of time until successful delivery is assumed to be geometric with mean $\mathbb{E}[S]=1/\mu$ slots, where $\mu$ is referred as the service probability.
In what follows we obtain the generating functions of PAoI and AoI utilizing the results of the previous section.

\begin{theorem}\label{theoremPreempt}
	The z-transforms of the PAoI and the AoI for the Geo/Geo/1 queue with an LCFS queue discipline with preemption are given by
	\begin{equation}
	A^{\ast}(z)	=\frac{\lambda \mu (\lambda (1-\mu)+\mu) z^2}{(1-(1-\lambda) z) (1-(1-\mu) z) (1-(1-\lambda) (1-\mu) z)},
	\label{eq:PAoI_gen_func_Pree}
	\end{equation}
	and
	\begin{equation}
	\Delta^{\ast}(z)= \frac{\lambda \mu z^2}{(1-(1-\lambda) z) (1-(1-\mu) z)},
	\label{eq:AoI_gen_func_Pree}
	\end{equation}	
	respectively.
	\begin{proof} The proof is given in Appendix~\ref{Appendix_D'}. \end{proof}	
\end{theorem}

As a result, taking the inverse z-transform we can obtain the pmfs of the PAoI and the AoI as
\begin{equation}
P_A(x) = \frac{(\lambda (1-\mu)+\mu) \left((\lambda-\mu) \left(\frac{1}{(1-\lambda) (1-\mu)}\right)^{1-x}-\lambda (1-\lambda)^{x-1}+\mu (1-\mu)^{x-1}\right)}{\lambda-\mu},
\label{eq:A_pmf_Pree}
\end{equation}
and
\begin{align}
&P_{\Delta}(x) = 
\frac{\lambda \mu \left((1-\lambda)^{x-1}-(1-\mu)^{x-1}\right)}{\mu-\lambda}.
\label{eq:Delta_pmf_Pree}
\end{align}

Furthermore, we derive the PDFs of the PAoI and the AoI as follows
\begin{align}
&A(x) =  \frac{1}{\lambda-\mu} \Bigg(
(\mu-\lambda) \left(\frac{1}{(\lambda-1) (\mu-1)}\right)^{-x}-\lambda (1-\mu)^x+\mu (1-\lambda)^x+\lambda \mu (1-\mu)^x \notag \\ 
& -\lambda \mu (1-\lambda)^x+\lambda (1-\lambda)^x+\lambda-\mu (1-\mu)^x-\mu \Bigg), 
\label{eq:A_PDF_Pree}
\end{align}
and
\begin{equation}
\Delta(x) = 
\frac{\lambda-\lambda (1-\mu)^x+\mu \left((1-\lambda)^x-1\right)}{\lambda-\mu}.
\label{eq:Delta_PDF_Pree}
\end{equation}
These can be utilized for quantifying the AoI or PAoI violation probability that can in turn provide PAoI and AoI performance guarantees for the system.

\section{Bufferless System with Packet Dropping}
Consider a system where the source generates a status update with average probability $\lambda$ and transmits it to the destination through an erasure channel with a success probability $p$. 
A packet that is not successfully transmitted is dropped from the system. 
Therefore, two consecutive successful receptions might result from multiple attempted transmissions, the number of which we denote by $M$.
We assume that packets arrive at the beginning of the time slot and finish service at the end of
the time slot.
Let $Y_i=t_i-t_{i-1}$ denote the time between two consecutive attempted transmissions.
Moreover, let $X_n=t'_{n}-t'_{n-1}$ be the random variable denoting the time between the reception of status update $n-1$ and $n$.
Then, we have that $X_n = \sum_{i=1}^{M} Y_i$.

To derive the z-transforms of the PAoI and AoI we proceed as follows.
We observe that the PAoI of update $n$ in this system coincides with the inter-reception times $X_n$, that is $A_n=X_n$.
%\begin{equation}
%A_n=X_n.
%\end{equation}
We note that $Y_i$, $i = 1,2, .. $ are independent and identically distributed geometric random variables with parameter $\lambda$.
Then, we condition on $M=m$ which occurs with probability $(1-p)^{m-1} p$ and obtain 
\begin{align}
A^{\ast}(z) &= \sum_{m=1}^{\infty} \left( \frac{\lambda z}{1-(1-\lambda)z}\right) ^m (1-p)^{m-1} p = \frac{\lambda p z}{1-(1-\lambda p)z}.
\label{eq:X_z_transform}
\end{align}
This implies that the pmf of the PAoI is given by 
\begin{equation}
P_A(x) = \lambda p (1-\lambda p)^{x-1}, \quad x>0.
\label{eq:PAoI_pmf_nobuffer}
\end{equation}
Therefore, the PDF of the PAoI is 
\begin{equation}
A(x) =\text{Pr} (A \leq x)=1-(1-\lambda p)^x.
\label{eq:PAoI_PDF_nobuffer}
\end{equation}

Next, we invoke Theorem~\ref{theorem2} and consider only the packets that are successfully transmitted.
We refer to the average arrival rate of packets that enter and
remain in the system as the effective arrival rate and define it as $\lambda_e = \lambda (1-p_D) = \lambda p$,
%\begin{equation}
%\lambda_e = \lambda (1-p_D) = \lambda p.
%\label{eq:effective_lambda}
%\end{equation}
where $p_D=1-p$ is the packet dropping probability.
The delay for a packet that is not dropped is one time slot.
As a result, the z-transform of the AoI is given by 
\begin{equation}
\Delta^{\ast}(z)= \lambda_e \frac{1-A^*(z)}{(1-z)/z} = \frac{\lambda p z}{1-(1-\lambda p)z}.
\label{eq:AoI_z_transform_no_buffer}
\end{equation}
We note that \eqref{eq:AoI_z_transform_no_buffer} is equal to \eqref{eq:X_z_transform}. 
This means that the PAoI and the AoI for the bufferless system with packet dropping have the same distribution.
As a result, the pmf of the AoI is given by \eqref{eq:PAoI_pmf_nobuffer}
and the PDF of the AoI is given by \eqref{eq:PAoI_PDF_nobuffer}.

\section{Non-linear Age Functions}

Besides the transmission delay, that is determined by system considerations such as the arrival process, the queueing model, and the service process, the AoI is affected by the frequency of the status update transmissions.
This sampling frequency can further be evaluated based on the characteristics of the source that indicate the cost of an update delay depending on the application.
Next, we consider a non-negative and monotonically increasing class of functions $f_s(t)$, $t\geq0$, with $f_s(0)=0$.
We define the cost of update delay (CoUD) to be
\begin{equation}
C_t = f_s (\Delta_t)= f_s(t-u(t)),
\end{equation}	
and the peak cost of update delay (PCoUD) as
\begin{equation}
C_{peak,n} = f_s (A_n)= f_s(t'_n-t_{n-1}).
\end{equation}	
In what follows, we provide a general formula that computes the time average CoUD and PCoUD and we derive closed-form expressions for various cases of the $f_s(t)$ cost function.
The method introduced for the average CoUD and PCoUD can be extended to the analysis of higher order moments and properties of the distributions of age processes.

\begin{theorem}\label{theorem5}
	For any	non-negative, non-decreasing function of age, $f_s(t)$, $t\geq0$, the time average PCoUD for a discrete time single-server queue is obtained as
	\begin{equation}
	C_{peak} = \lim_{\mathcal{T} \to \infty} \frac{1}{\mathcal{T}}\sum_{n=1}^{N(\mathcal{T})} C_{peak,n} = \sum_{n=1}^{\infty} f_s(n) P_{A}(n),
	\label{eq:average_peak_age}
	\end{equation}
	and the time average CoUD is obtained as
	\begin{equation}
	C = \lim_{\mathcal{T} \to \infty} \frac{1}{\mathcal{T}}\sum_{t=0}^{\mathcal{T}} C_t = \sum_{t=0}^{\infty} f_s(t) P_{\Delta}(t).
	\label{eq:average_age}
	\end{equation}
	\begin{proof} The relations in \eqref{eq:average_peak_age} and \eqref{eq:average_age} immediately follow from the definition of time average and the fundamental theorem of expectation \cite[Page 379]{Kleinrock}.
	\end{proof}			
\end{theorem}

\subsection{Time Average Analysis: Three examples}
\subsubsection{The FCFS Geo/Geo/1 queue}
For this queue discipline we utilize Theorem~\ref{theorem5} together with the pmfs of the PAoI and AoI in \eqref{eq:A_pmf} and \eqref{eq:Delta_pmf}, respectively.

For $f_s(t)=\alpha t$, $\alpha >0$, the average CoUD for the Geo/Geo/1 system with an FCFS queue discipline is given by
\begin{equation}
C= \alpha \left( \frac{1}{\lambda}+\frac{1-\lambda}{ \mu-\lambda}-\frac{\lambda}{\mu^2}+\frac{\lambda}{\mu} \right),
\label{eq:average_age_lin}
\end{equation}	
and the average PCoUD is given by
\begin{equation}
C_{peak} = \alpha \frac{\lambda^2-\mu}{\lambda (\lambda-\mu)}.
\label{eq:average_peakage_lin}
\end{equation}	
For $\alpha= 1$, the results in \eqref{eq:average_age_lin} and \eqref{eq:average_peakage_lin} are the AoI and PAoI of an Geo/Geo/1 queue that are found in \cite{Kosta19_ISIT}, respectively.

For $f_s(t)=\alpha t^2$, $\alpha >0$, the average CoUD for the Geo/Geo/1 system with an FCFS
queue discipline is given by
\begin{align}
C &= \frac{\alpha}{\lambda^2 \mu^3 (\lambda-\mu)^2} \big( \lambda^5 (\mu-1) (\mu+4)+\lambda^4 \mu ((\mu-8) \mu+8)
\notag \\ 
&+\lambda^3 (\mu-2) \mu^2+\lambda^2 \mu^4-\lambda \mu^4 (\mu+2)+2 \mu^5 \big),
\label{eq:average_age_pow}
\end{align}	
and the average PCoUD is given by
\begin{align}
C_{peak} &= \frac{\alpha}{\lambda^2 \mu^2 (\mu-\lambda)^2} \big( \lambda^4 (\mu (\mu+2)-2) +\lambda^3 \mu ((\mu-6) \mu \notag \\ 
& +4)+\lambda^2 \mu^3-\lambda \mu^3 (\mu+2)+2 \mu^4 \big).
\label{eq:average_peakage_pow}
\end{align}	

Next, we consider a general exponentiation to a non-negative integer power $n$. This result
will prove to be useful for representing functions as power series.
For $f_s(t)=\alpha t^n$, $\alpha >0$, the average CoUD for the Geo/Geo/1 system with an FCFS queue discipline is given by
\begin{align}
C &= \frac{\alpha}{\lambda-\mu} \Bigg( \frac{\lambda \mu \Phi (1-\lambda,-n,0)}{\lambda-1} + \frac{1}{(1-\mu)^2} \bigg( \left( \lambda^2+(1-2 \lambda) \mu^2+(\lambda-1) \lambda \mu \right) \Phi (1-\mu,-n,0)+ \notag \\ 
& +(\lambda-1) (\mu-\lambda)^2 \Phi \left(\rho,-n,0\right)+\lambda \mu (\mu-\lambda) \Phi (1-\mu,-n-1,0) \bigg) \Bigg),
\label{eq:average_age_pow_n}
\end{align}
where $\Phi$ is the Lerch transcendent (Hurwitz-Lerch zeta function) defined as
%Lerch transcendent function 
%Hurwitz-Lerch zeta-function 
\begin{equation}
\Phi(z,s,\beta) = \sum_{\eta=0}^{\infty} \frac{z^\eta}{(\eta+\beta)^s}, \quad \arrowvert z \arrowvert<1, \quad \beta \neq 0, -1,\cdots,
\label{eq:Phi_function}
\end{equation}		
and any term with $n+\beta=0$ is excluded.

Furthermore, the average PCoUD is given by
\begin{align}
C_{peak} &= \frac{\alpha \mu}{\lambda (\lambda-\mu)} \Bigg( \frac{\lambda^2 \text{Li}_{-n}(1-\lambda)}{\lambda-1} + \frac{1}{(1-\mu)^2} \bigg(  \left(2 \lambda^2-\lambda \mu (\mu+2)+\mu^2\right) \text{Li}_{-n}(1-\mu) +\notag \\ 
&+(\lambda-1) (\mu-\lambda)^2 \text{Li}_{-n} \left(\rho\right)+\lambda \mu (\mu-\lambda) \text{Li}_{-n-1}(1-\mu) \bigg)  \Bigg),
\label{eq:average_peakage_pow_n}
\end{align}	
where $\text{Li}$ is the polylogarithm function defined as
\begin{equation}
z \Phi(z, s, 1) = \text{Li}_{s}(z) \doteq \sum_{\eta=1}^{\infty} \frac{z^\eta}{\eta^s}, \quad \arrowvert z \arrowvert<1, \quad s \in \mathbb{C}. 
\label{eq:polylogarithm_function}
\end{equation}	

\subsubsection{The Preemptive LCFS Geo/Geo/1 queue}
For this queue discipline we utilize Theorem~\ref{theorem5} together with the pmfs of the PAoI and AoI in \eqref{eq:A_pmf_Pree} and \eqref{eq:Delta_pmf_Pree}, respectively.

For $f_s(t)=\alpha t$, $\alpha >0$, the average CoUD for the preemptive Geo/Geo/1 system with an LCFS queue discipline is given by
\begin{equation}
C= \alpha \left(\frac{1}{\lambda}+\frac{1}{\mu}\right),
\label{eq:average_age_lin_Pree}
\end{equation}	
and the average PCoUD is given by
\begin{equation}
C_{peak} = \frac{\alpha \left(\lambda^2 (1-\mu)^2+\lambda \mu (3-2 \mu)+\mu^2\right)}{\lambda \mu (\lambda (1-\mu)+\mu)}.
\label{eq:average_peakage_lin_Pree}
\end{equation}	

For $f_s(t)=\alpha t^2$, $\alpha >0$, the average CoUD for the preemptive Geo/Geo/1 system with an LCFS
queue discipline is given by
\begin{equation}
C = \frac{a \left(\lambda^2 (2-\mu)+\lambda (2-\mu) \mu+2 \mu^2\right)}{\lambda^2 \mu^2},
\label{eq:average_age_pow_Pree}
\end{equation}	
and the average PCoUD is given by
\begin{equation}
C_{peak} = \alpha \left(\frac{2}{\lambda^2}+\frac{\frac{4}{\mu}-3}{\lambda}-\frac{1}{\lambda+\mu-\lambda \mu}+\frac{2}{(\lambda+\mu-\lambda \mu)^2}+\frac{2-3 \mu}{\mu^2}+1\right).
\label{eq:average_peakage_pow_Pree}
\end{equation}	

For $f_s(t)=\alpha t^n$, $\alpha >0$, the average CoUD for the preemptive Geo/Geo/1 system with an LCFS queue discipline is given by
\begin{equation}
C = \frac{\alpha \lambda \mu ((\mu-1) \Phi (1-\lambda,-n,0)-(\lambda-1) \Phi (1-\mu,-n,0))}{(1-\lambda) (1-\mu) (\lambda-\mu)},
\label{eq:average_age_pow_n_Pree}
\end{equation}
where $\Phi$ is the Lerch transcendent (Hurwitz-Lerch zeta function) defined in \eqref{eq:Phi_function}.

Furthermore, the average PCoUD is given by
\begin{align}
C_{peak} &= \frac{\alpha (\lambda (\mu-1)-\mu)}{(1-\lambda) (1-\mu) (\mu-\lambda)} \Big( \lambda (\mu-1) \text{Li}_{-n} (1-\lambda) +(\mu-\lambda \mu) \text{Li}_{-n} (1-\mu) \notag \\ 
&+(\lambda-\mu) \text{Li}_{-n} \big( (\lambda-1) (\mu-1) \big) \Big),
\label{eq:average_peakage_pow_n_Pree}
\end{align}	
where $\text{Li}$ is the polylogarithm function defined in \eqref{eq:polylogarithm_function}.

\subsubsection{The Bufferless System with Packet Dropping}
For this queue discipline we utilize Theorem~\ref{theorem5} together with the pmfs of the PAoI and AoI in \eqref{eq:PAoI_pmf_nobuffer}.

For $f_s(t)=\alpha t$, $\alpha >0$, the average CoUD and PCoUD for the bufferless system with packet dropping are given by
\begin{equation}
C = C_{peak}  =  \frac{\alpha}{\lambda p}.
\label{eq:average_age_lin_noQueue}
\end{equation}	

For $f_s(t)=\alpha t^2$, $\alpha >0$, the average CoUD and PCoUD for the bufferless system with packet dropping are given by
\begin{equation}
C = C_{peak} = \frac{\alpha (2-\lambda p)}{\lambda^2 p^2}.
\label{eq:average_age_pow_noQueue}
\end{equation}	

For $f_s(t)=\alpha t^n$, $\alpha >0$, the average CoUD for the bufferless system with packet dropping is given by
\begin{equation}
C = \frac{\alpha \lambda p \Phi (1-\lambda p,-n,0)}{1-\lambda p},
\label{eq:average_age_pow_n_noQueue}
\end{equation}
and the average PCoUD is given by
\begin{equation}
C_{peak} = \frac{\alpha \lambda p \text{Li}_{-n}(1-\lambda p)}{1-\lambda p}.
\label{eq:average_peakage_pow_n_noQueue}
\end{equation}

\subsection{Representation of functions as power series}

The result in Theorem~\ref{theorem5} provides a convenient way to determine the time average PCoUD and CoUD by representing functions with power series.
This observation is formalized in the following claim, which is a restatement of Theorem~\ref{theorem5} in compact form.
For any non-negative, non-decreasing function of age $f_s(t)$, we can derive the time average PCoUD and CoUD exact closed-form expressions, using a power series representation. 
A power series (in one variable) is an infinite series of the form
\begin{equation}
\sum_{k=0}^{\infty}  \omega_k (x-\xi)^k = \omega_0 +  \omega_1 (x-\xi)^1 + \omega_2 (x-\xi)^2 + \cdots 
\label{eq:power_series}
\end{equation}	
where $\omega_k$ represents the coefficient of the $k$th term and $\xi$ is a constant.
The following is true for any power series: if it is not everywhere convergent, the region of convergence is a circle with its center at the point $\xi$ and a radius equal to $R$; at every interior
point of this circle, the power series converges absolutely, and outside this circle, it diverges. 
One case is Taylor's theorem that gives an approximation of a $k$-times differentiable function around a given point by a $k$th order Taylor polynomial.
To provide a methodology for mean analysis of general non-linear age functions, we demonstrate how to use series expansion to obtain the average CoUD and PCoUD of any non-negative and monotonically increasing class of functions, provided that the series converges.

\LinesNumberedHidden{
	\begin{algorithm}
		\normalsize    
		\caption{Time Average Performance Analysis} \label{algo:TAP_algo}
		\textbf{Step 1:} Initialize $f_s(t)$, $\alpha$, $t$, and $GAP$. 
		~~Set the number of terms $k>0$. Give termination criteria $\epsilon$. \\
		
		\textbf{Step 2:} Determine $f_{approx}(k)$ according to  Sec. IX. B.	~~Set $ GAP := \left\lVert f_s - f_{approx}(k) \right\rVert$.\\

		\textbf{Step 3:} If $GAP < \epsilon$, then go to step 5. \\
		
		\textbf{Step 4:} Update $f_{approx}(k)$ by adding an extra term;
		set $ k := k + 1 $. Go to step $2$. \\
		
		\textbf{Step 5:} Obtain the time average expressions according to Theorem~\ref{theorem5}. Return $C$ and $C_{peak}$ as the solution.		
	\end{algorithm}   	
}

The proposed Algorithm 1 provides guidelines for the time average performance analysis of functions that are not in a polynomial form.
%In step 1, 
The algorithm offers control parameters, e.g., $GAP$ and $\epsilon$, that can be used to trade the accuracy of obtaining the exact closed-form expression of the time average CoUD and PCoUD. 

Example 1: Consider the exponential function $f_s(t)=e^{\alpha t}-1$.
The Taylor expansion of this function evaluated at 0, is given by 
\begin{equation}
e^{\alpha t}-1 = \sum_{k=0}^{\infty} \frac{(\alpha t)^k}{k!} -1 = \alpha t+\frac{ \alpha^2 t^2}{2}+\frac{ \alpha^3 t^3}{6}+\frac{ \alpha^4 t^4}{24}+\dots. 
\label{eq:exp_Taylor_expan}
\end{equation}	
We fix $t$ and determine the approximation
%\begin{equation}
%f_{approx}(k_{max})=\sum_{k=0}^{k_{max}} \frac{(\alpha t)^k}{k!} -1. 
%\label{eq:exp_f_approx}
%\end{equation}	
\begin{equation}
f_{approx}(k)=f_{approx}(k-1) + \frac{(\alpha t)^k}{k!}, \quad k>0,
\label{eq:exp_f_approx}
\end{equation}	
where $f_{approx}(0)=0$.
We notice that the expression in \eqref{eq:exp_f_approx} consists of terms of the form $\alpha t^n$ that we analyzed in the previous subsection.

Example 2: Consider the logarithmic function $f_s(t)= \log(\alpha t+1)$.
A series representation of this function, is given by 
\begin{equation}
\log(\alpha t+1) = 2 \sum_{k=1}^{\infty} \frac{1}{2k-1} \left( \frac{\alpha t}{\alpha t+2}\right)^{2k-1} = 2 \left( \frac{\alpha t }{\alpha t +2} + \frac{1}{3} \left( \frac{\alpha t}{\alpha t +2} \right)^3 + \frac{1}{5} \left( \frac{\alpha t}{\alpha t +2} \right)^5 + \dots \right).
\label{eq:log_expan}
\end{equation}
We fix $t$ and determine the approximation
\begin{equation}
f_{approx}(k) = f_{approx}(k-1) + 2 \frac{1}{2k-1} \left( \frac{\alpha t}{\alpha t+2}\right)^{2k-1}, \quad k>1, 
\label{eq:log_f_approx}
\end{equation}
where $f_{approx}(0)=0$.		
The equations in \eqref{eq:exp_f_approx} and \eqref{eq:log_f_approx} can be used iteratively as $f_{approx}(k)$ in Algorithm 1.
In Section~\ref{sec:NumericalResults}, we illustrate how the number of terms $k$ affects the CoUD and PCoUD performance analysis.

\subsection{Optimal Arrival Probability for the FCFS Geo/Geo/1 Queue}

In order to find the optimal value of $\lambda$ that minimizes the average CoUD and PCoUD, for a given $\mu$, we proceed as follows.
We differentiate $C$ and $C_{peak}$ with respect to $\lambda$ to obtain $\frac{\partial C}{\partial \lambda}$ and $\frac{\partial C_{peak}}{\partial \lambda}$, respectively.
Then, for both CoUD and PCoUD, taking the second derivative $\frac{\partial^2 C}{\partial \lambda^2}$ and $\frac{\partial^2 C_{peak}}{\partial \lambda^2}$ it can be easily seen that $C$ and $C_{peak}$ are convex functions of $\lambda$ for a given service rate $\mu$, if $\lambda<\mu$ is not violated.

Case 1: For \underline{$f_s(t)=\alpha t$}, by setting $\frac{\partial C}{\partial \lambda}=0$ we can obtain the value of $\lambda$ that minimizes the CoUD and satisfies the equation $\alpha (\lambda^4 (\mu-1) - 2 \lambda^3 (\mu-1) \mu- \lambda^2 \mu^2 +2 \lambda \mu^3 -\mu^4) = 0$.
Moreover, by setting $\frac{\partial C_{peak}}{\partial \lambda}=0$ we can obtain the value of $\lambda$ that minimizes the PCoUD and satisfies the equation $\alpha \mu ((\lambda-2) \lambda+\mu) = 0$.
For PCoUD, the resulting optimal arrival probability is $\lambda^{\ast}=1-\sqrt{1-\mu}$.

Case 2: For $\underline{f_s(t)=\alpha t^2}$, by setting $\frac{\partial C}{\partial \lambda}=0$ we can obtain the value of $\lambda$ that minimizes the CoUD and satisfies the equation $\alpha (\lambda^6 (\mu-1) (\mu+4)-3 \lambda^5 (\mu-1) \mu (\mu+4)+\lambda^4 \mu^2 ((15-2 \mu) \mu-14)+\lambda^3 \mu^3 (2-3 \mu)+3 \lambda^2 \mu^4 (\mu+2)-\lambda \mu^5 (\mu+10)+4 \mu^6)= 0$.
Moreover, by setting $\frac{\partial C_{peak}}{\partial \lambda}=0$ we can obtain the value of $\lambda$ that minimizes the PCoUD and satisfies the equation $a ((\lambda-2) \lambda+\mu) ( \lambda^2 (3 \mu-2)+\lambda \mu (\mu+2)-4 \mu^2 ) = 0$.
For PCoUD, the resulting optimal arrival probability is $\lambda^{\ast}=1-\sqrt{1-\mu}$.

Case 3: For $\underline{f_s(t)=\alpha t^2 +\beta t}$, $\beta>0$, working similarly to the previous cases, by setting $\frac{\partial C_{peak}}{\partial \lambda}=0$ and utilizing elementary algebra that we skip due to the lengthy expression, we can obtain the value of $\lambda$ that minimizes the PCoUD to be $\lambda^{\ast}=1-\sqrt{1-\mu}$.

\begin{remark}
	An interesting observation is that in case of PCoUD all functions are minimized by the same value of $\lambda^*$ and independently of the choice of the parameter $\alpha$.
	Furthermore, for both CoUD and PCoUD we note that the optimal $\lambda$ is independent of $\alpha$, that is a scaling factor in the equations of the derivative.
\end{remark}

\begin{proposition}
	Consider the average PCoUD in the FCFS Geo/Geo/1 queue.
	The optimal arrival probability $\lambda^*$ that minimizes PCoUD is the same for every non-negative and monotonically increasing polynomial function $f_s(t)$, $t\geq0$, with $f_s(0)=0$.
	\begin{proof} 
		%The proof follows  
		To obtain the optimal arrival probability $\lambda^{\ast}$ we follow the approach illustrated in Cases 1, 2, and 3, for any non-negative and monotonically increasing polynomial function $f_s(t)$.
		By setting $\frac{\partial C_{peak}}{\partial \lambda}=0$ and utilizing elementary algebra we result to the optimal arrival probability $\lambda^{\ast}=1-\sqrt{1-\mu}$	which is independent of $\alpha$.
		Hence, $\lambda^*$ is identical for all such functions $f_s(t)$.
	\end{proof}	
\end{proposition}

\vspace*{-0.9cm}
\section{Numerical Results}	\label{sec:NumericalResults}
In this section, we evaluate the performance of information freshness in the considered system models using numerical results.
In addition, we develop a MATLAB-based behavioral simulator where each case runs for $10^6$ time slots, to validate the analytical results.

\subsection{The stationary distribution of the AoI}
\begin{figure*}[t!]
	\centering
	%\hspace*{-0.1in}
	\subfloat[][]{
		\includegraphics[draft=false,scale=.45]{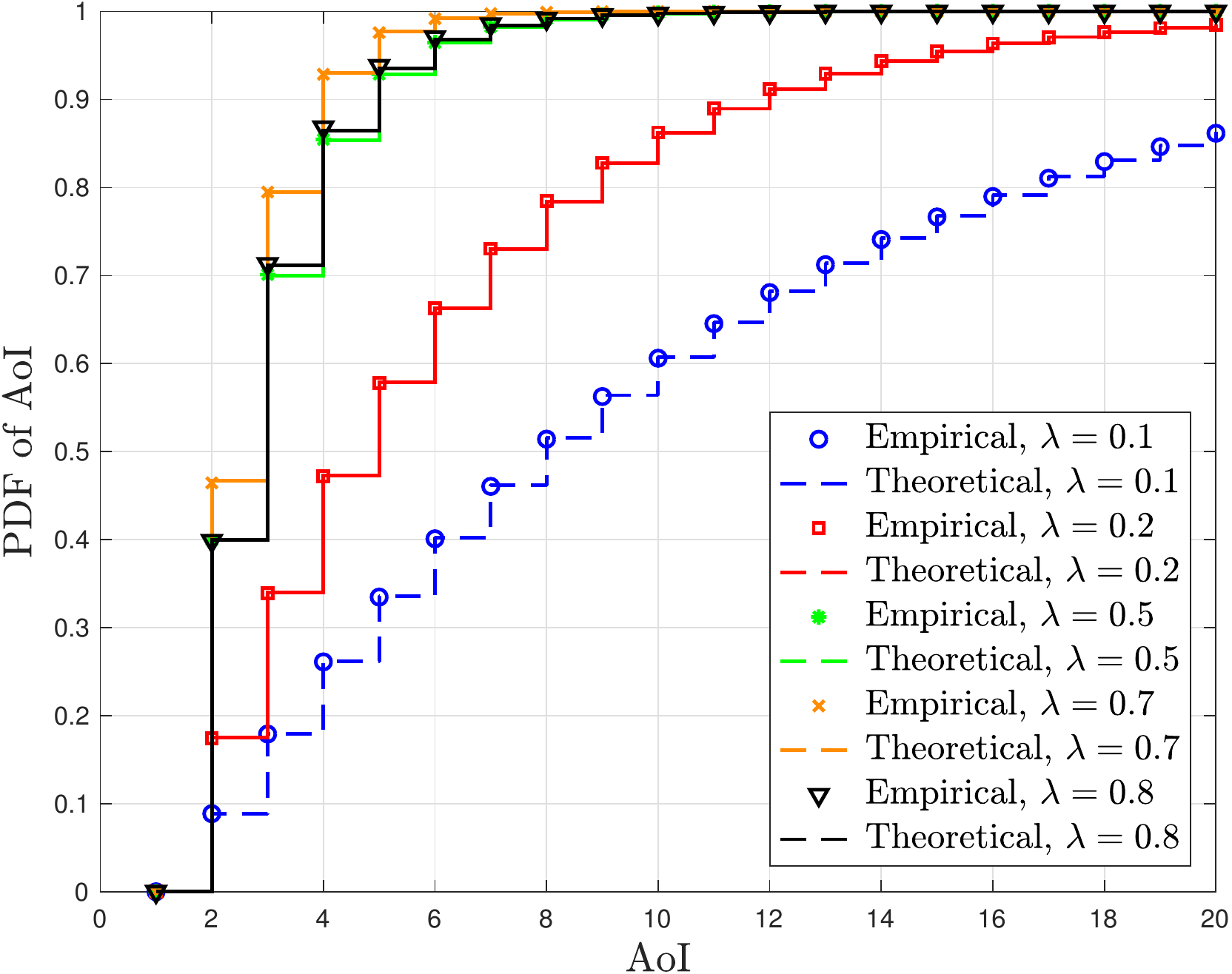}
		\label{fig:CDF_of_AoI_geo_geo_1}
	}
	\subfloat[][]{
		%\rule{0.3\linewidth}{3cm}
        \includegraphics[draft=false,scale=.45]{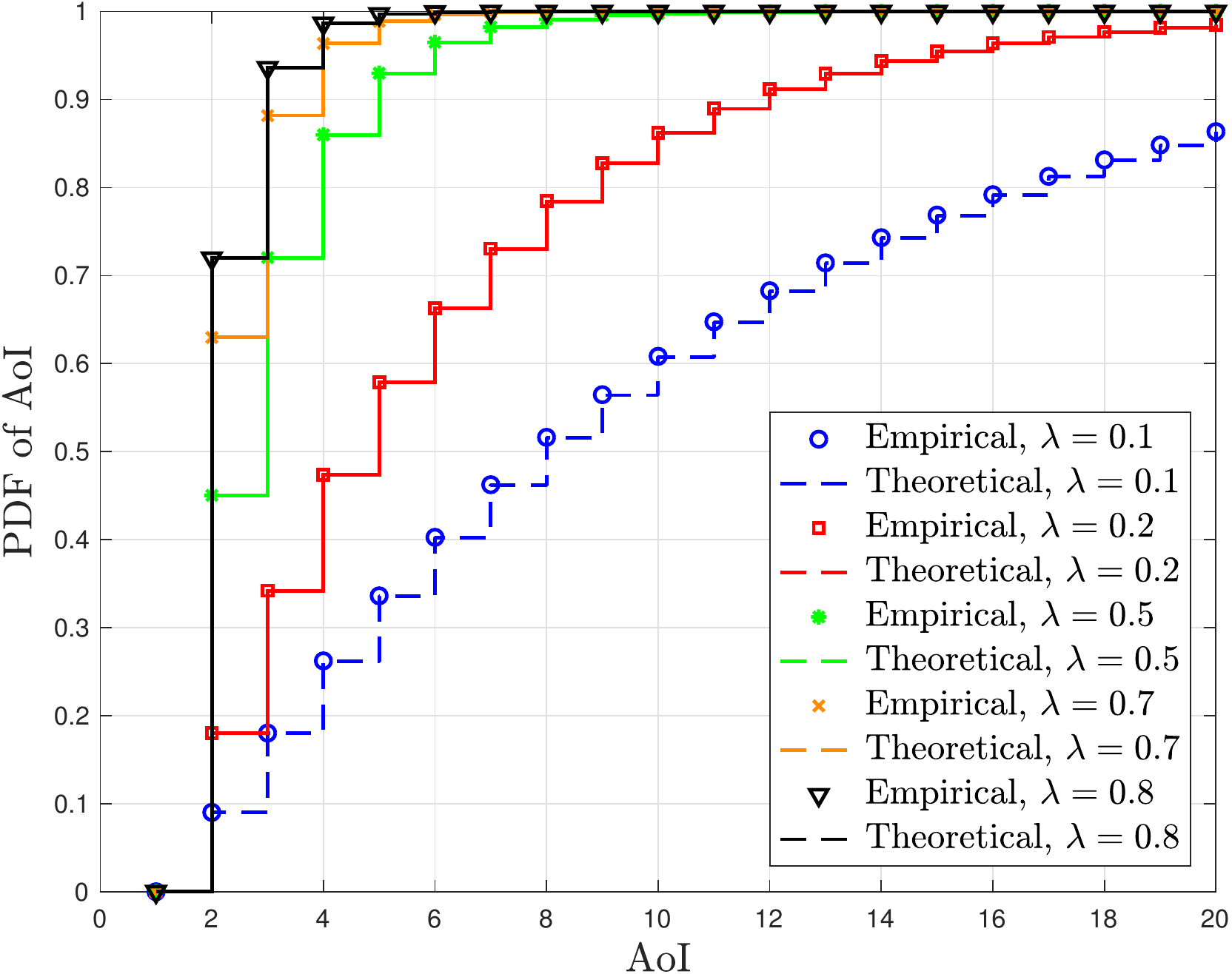}
		\label{fig:CDF_of_AoI_preemptiveLCFS}
	}
	%\vspace{-3mm}
	\caption{The stationary distribution of the AoI vs. the AoI, for the FCFS Geo/Geo/1 queue (left) and the preemptive LCFS Geo/Geo/1 queue (right) with $\mu=0.9$.}
	\label{fig:CDF_of_AoI_FCFS_ANDpreemptiveLCFS}
\end{figure*}

In Fig.~\ref{fig:CDF_of_AoI_geo_geo_1}, we depict the PDF of the AoI in \eqref{eq:Delta_PDF} for the FCFS Geo/Geo/1 queue, as a function of the AoI, for different values of $\lambda$, and $\mu=0.9$. 
We observe that as the average probability of arrival $\lambda$ increases, the probability of the AoI being less than or equal to a given value increases as well.
However, AoI is not a monotonically decreasing function of $\lambda$. 
For $\lambda=0.8$, when $\lambda$ approaches the service rate $\mu$, and the queue tends to become unstable, AoI increases.
These results can be utilized to provide AoI performance guarantees in terms of the AoI violation probability.

In Fig.~\ref{fig:CDF_of_AoI_preemptiveLCFS}, we depict the PDF of the AoI in \eqref{eq:Delta_PDF_Pree} for the preemptive LCFS Geo/Geo/1 queue as a function of the AoI, for different values of $\lambda$, and $\mu=0.9$. 
For this discipline, we observe that AoI is a monotonically decreasing function of $\lambda$. 
This means that the arrival probability that minimizes the AoI tends to the service probability $\mu$. 
Overall, the preemptive LCFS queue performs better than the FCFS queue in terms of the stationary distribution of AoI, for all $\lambda$.

\subsection{The probability mass function of the AoI}
\begin{figure*}[t!]
	\centering
	%\hspace*{-0.1in}
	\subfloat[][]{
		%\rule{0.3\linewidth}{3cm}
		%\input{sawtooth_linear.tex}
		\includegraphics[draft=false,scale=.45]{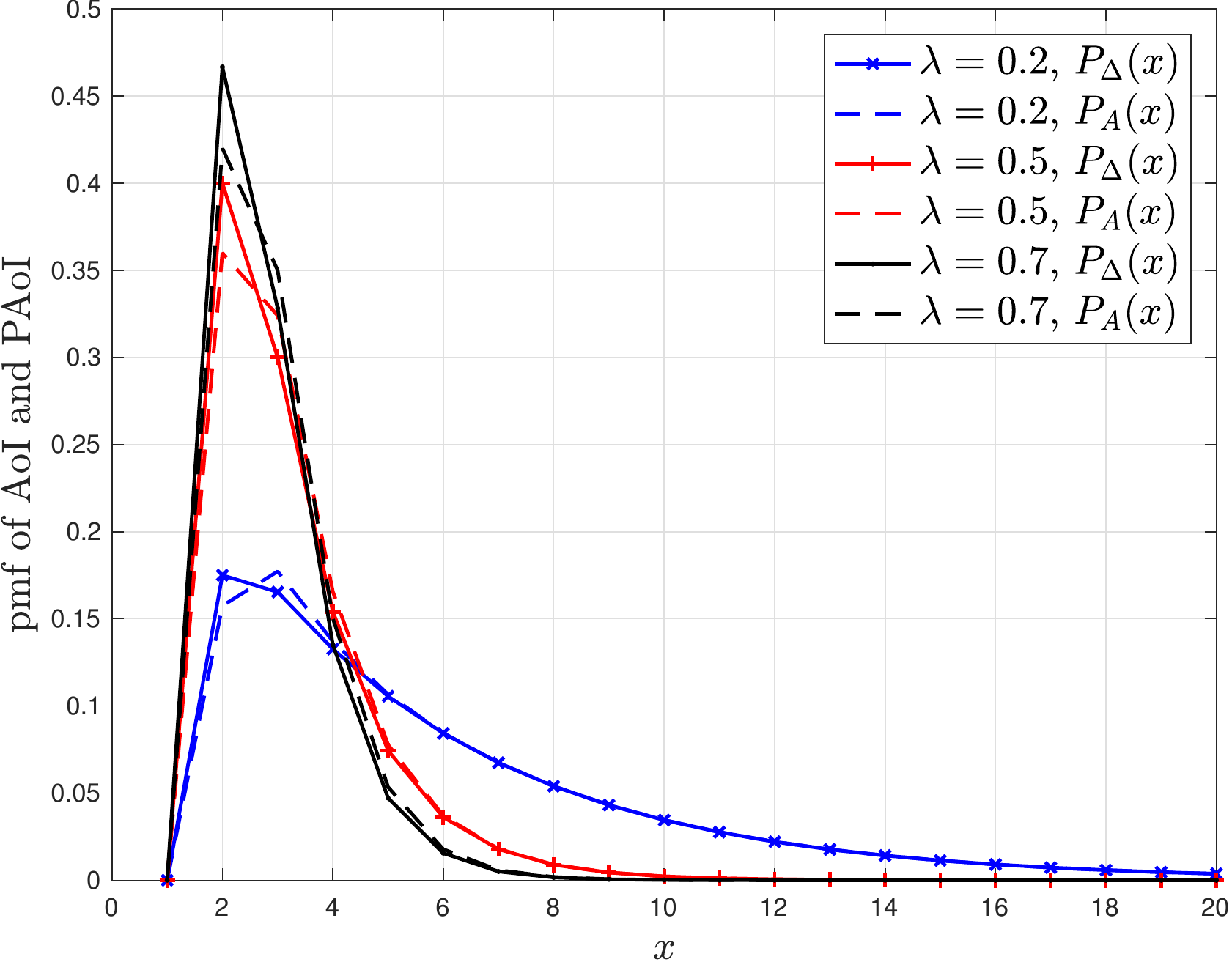}
		\label{fig:AoI_PAoI_pmf_vs_x_geogeo1}
	}
	\subfloat[][]{
		%\rule{0.3\linewidth}{3cm}
		\includegraphics[draft=false,scale=.45]{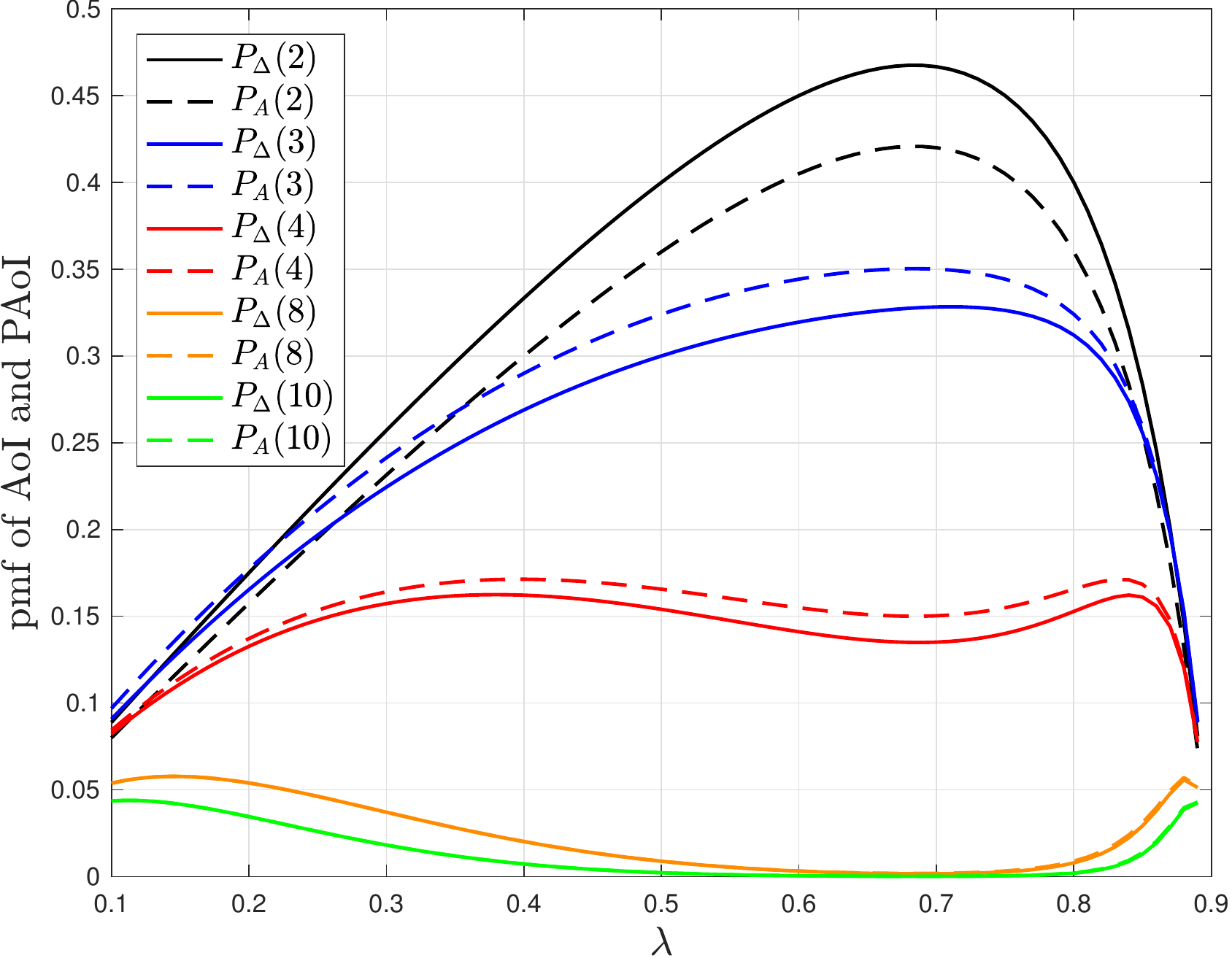}
		\label{fig:AoI_PAoI_pmf_vs_lambda_geogeo1}
	}
	%\vspace{-3mm}
	\caption{The pmfs of the AoI and the PAoI vs. $x$ (left) and the arrival probability $\lambda$ (right), for the FCFS Geo/Geo/1 queue with $\mu=0.9$.}
	\label{fig:AoI_PAoI_pmf_vs_xANDlambda_geogeo1}
\end{figure*}

\begin{figure*}[t!]
	\centering
	%\hspace*{-0.1in}
	\subfloat[][]{
		%\rule{0.3\linewidth}{3cm}
		\includegraphics[draft=false,scale=.45]{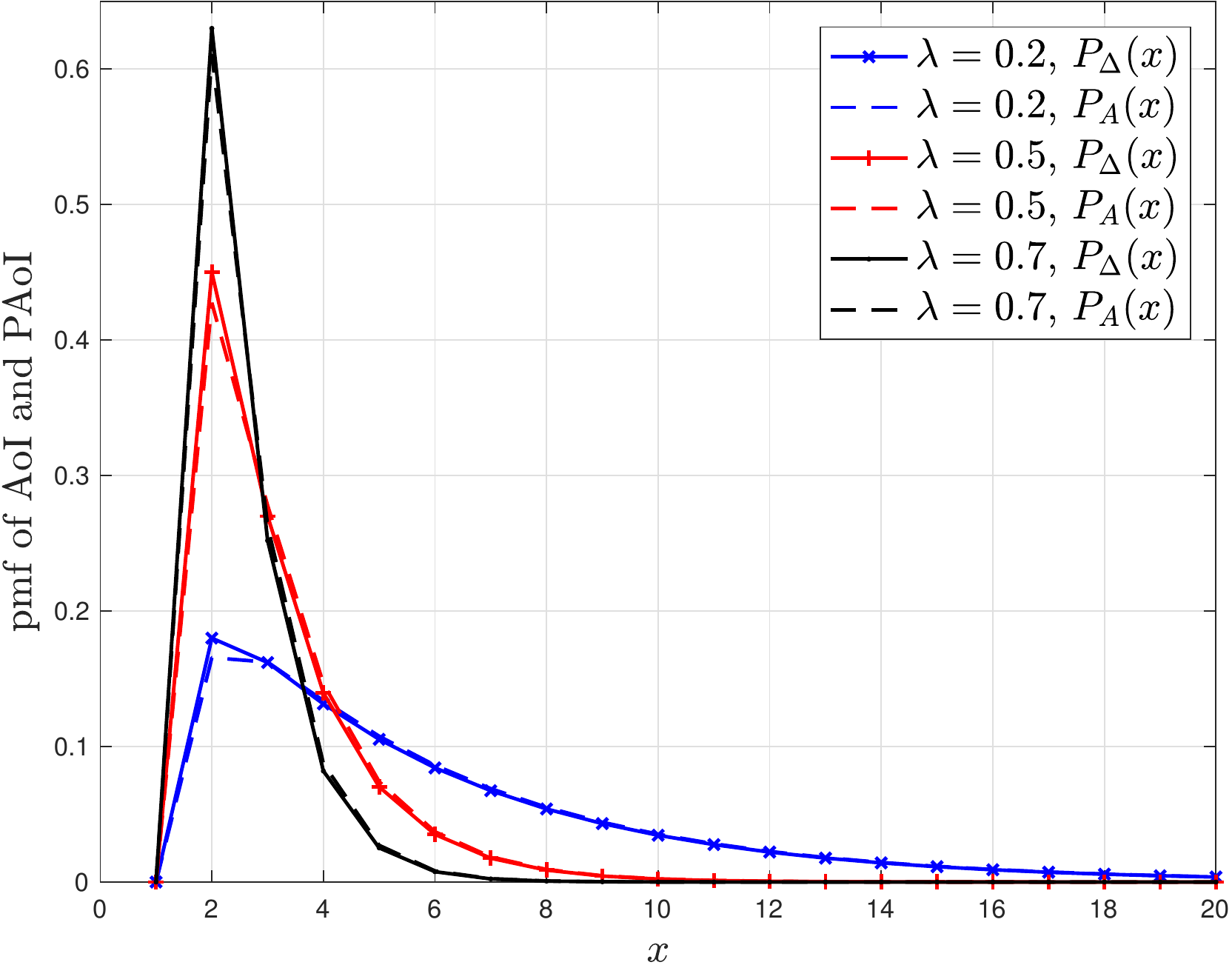}
		\label{fig:AoI_PAoI_pmf_vs_x_preemptiveLCFS}
	}
	\subfloat[][]{
		%\rule{0.3\linewidth}{3cm}
		\includegraphics[draft=false,scale=.45]{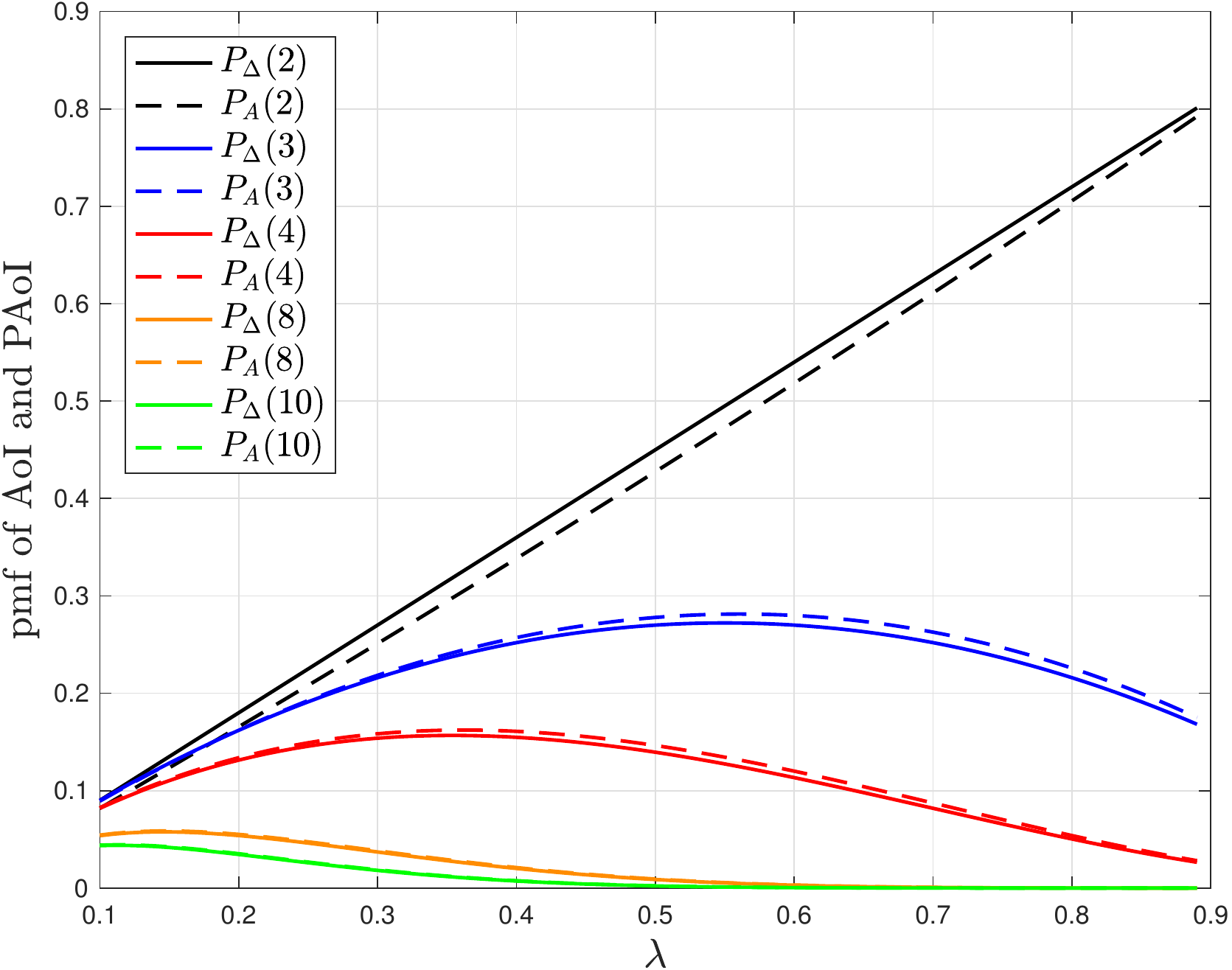}
		\label{fig:AoI_PAoI_pmf_vs_lambda_preemptiveLCFS}
	}
	%\vspace{-3mm}
	\caption{The pmfs of the AoI and the PAoI vs. $x$ (left) and the arrival probability $\lambda$ (right), for the preeemptive LCFS Geo/Geo/1 queue with $\mu=0.9$.}
	\label{fig:AoI_PAoI_pmf_vs_xANDlambda_preemptiveLCFS}
\end{figure*}

Next, we consider the probability that the AoI and the PAoI equal an arbitrary fixed value and compare the performance of the considered system models. 
In Fig.~\ref{fig:AoI_PAoI_pmf_vs_xANDlambda_geogeo1}, we plot the pmfs of the AoI and the PAoI in \eqref{eq:Delta_pmf} and \eqref{eq:A_pmf}, respectively, as a function of $x$ and $\lambda$. 
%To determine the optimal arrival probability $\lambda$ we should look at the different values of AoI and PAoI that the system requires.
In this case, the optimal arrival probability $\lambda^*$ can be determined based on a given target value $P_{\Delta}(x)$ and $P_{A}(x)$.
We note that both AoI and PAoI follow a similar behavior while for values greater that 4 the two pmfs almost coincide.
Furthermore, in Fig.~\ref{fig:AoI_PAoI_pmf_vs_xANDlambda_preemptiveLCFS} we plot the pmfs of the AoI and the PAoI in \eqref{eq:Delta_pmf_Pree} and \eqref{eq:A_pmf_Pree}, respectively, as a function of $x$ and $\lambda$. 
Notice that, due to preemption, to achieve the minimum AoI and PAoI, the arrival probability should be the highest possible.

\begin{figure*}[t!]
	\centering
	%\hspace*{-0.1in}
	\subfloat[][]{
		%\rule{0.3\linewidth}{3cm}
		\includegraphics[draft=false,scale=.45]{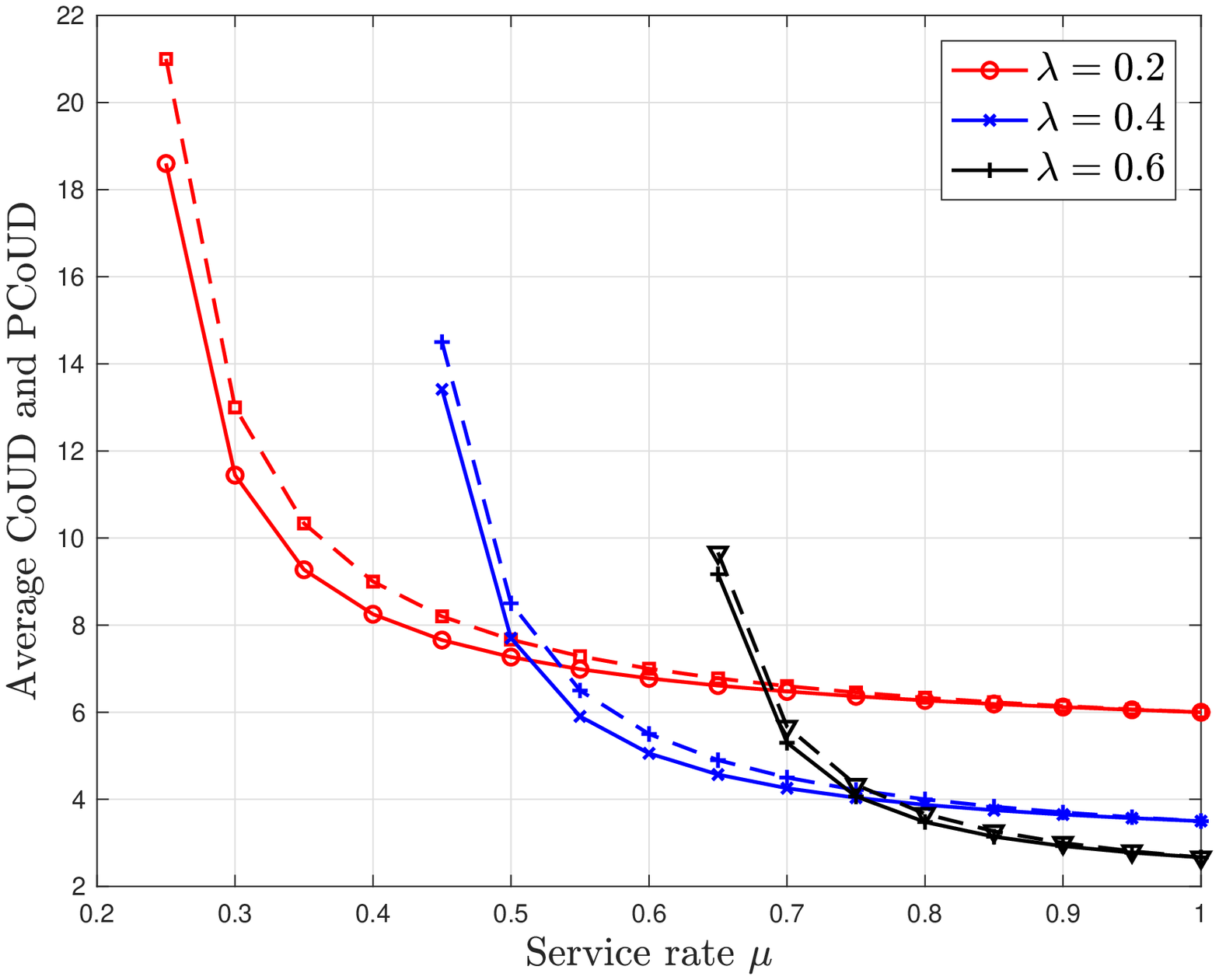}
		\label{fig:CoUD_PCoUD_vs_mu_geo_geo_1}
	}
	\subfloat[][]{
		%\rule{0.3\linewidth}{3cm}
		\includegraphics[draft=false,scale=.45]{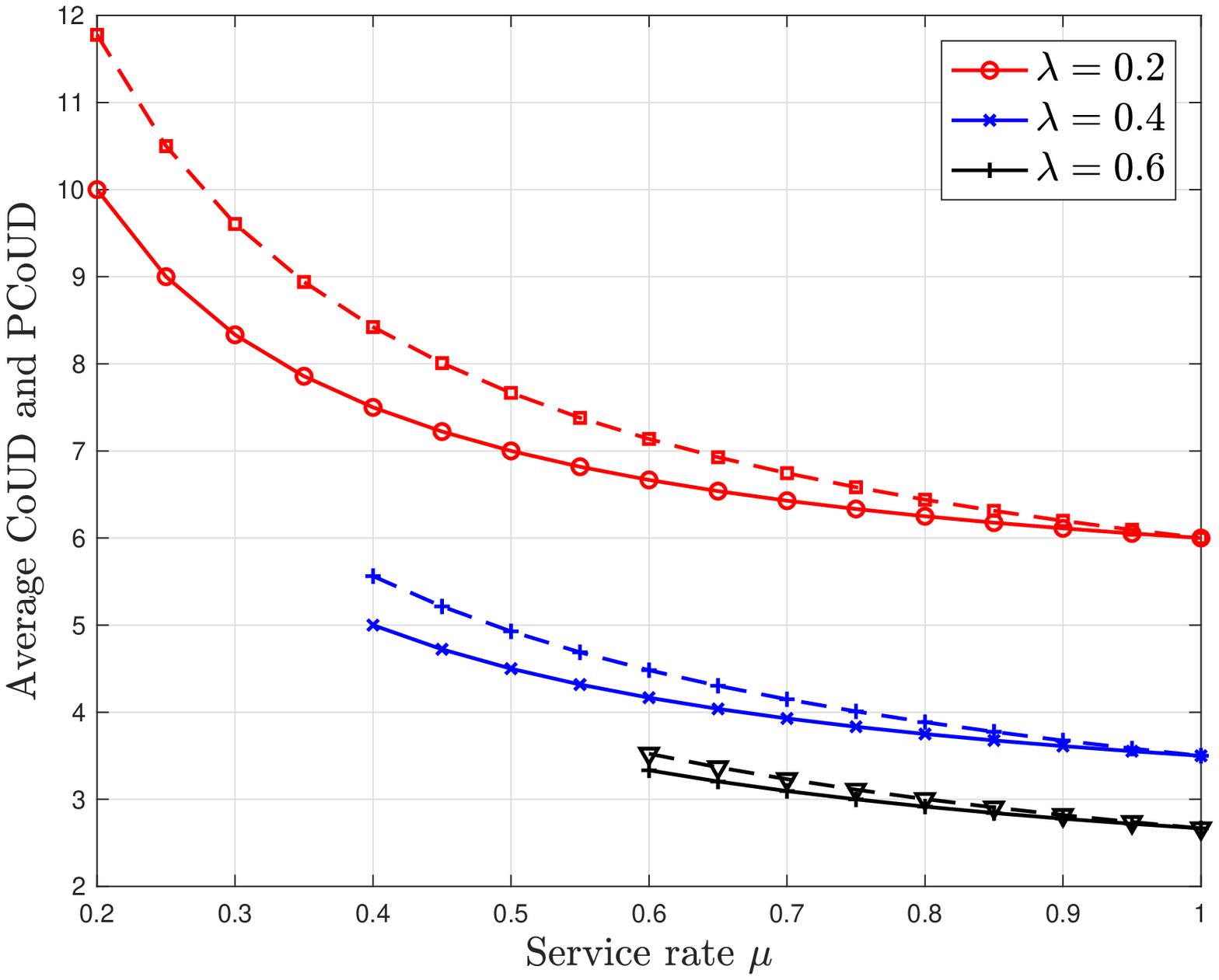}
		\label{fig:CoUD_PCoUD_vs_mu_preemptiveLCFS_geo_geo_1}
	}
	%\vspace{-3mm}
	\caption{The average CoUD and PCoUD for the linear case vs. the service probability $\mu$ for the FCFS Geo/Geo/1 queue (left) and the preemptive LCFS Geo/Geo/1 queue (right).}
	\label{fig:CoUD_PCoUD_vs_mu}
\end{figure*}

\subsection{The average CoUD and PCoUD}
We continue with the average performance analysis. 
Fig.~\ref{fig:CoUD_PCoUD_vs_mu} shows the time average CoUD (solid lines) and PCoUD (dashed lines) for $f_s(t)=t$, as a function of the service probability $\mu$.
In Fig.~\ref{fig:CoUD_PCoUD_vs_mu_geo_geo_1} we plot the results of the FCFS Geo/Geo/1 queue and in Fig.~\ref{fig:CoUD_PCoUD_vs_mu_preemptiveLCFS_geo_geo_1} we plot the results of the preemptive LCFS Geo/Geo/1 queue.
We observe that the PCoUD upper-bounds the CoUD and that the gap is large when $\mu$ is close to $\lambda$, and diminishes as $\mu$ increases.
%Large $\mu$ implies less ''interaction'' among packets.
Moreover we notice that the preemptive LCFS Geo/Geo/1 queue outperforms the FCFS Geo/Geo/1 queue. 

\begin{figure}[t!]\centering
	\centering
	\includegraphics[draft=false,scale=.5]{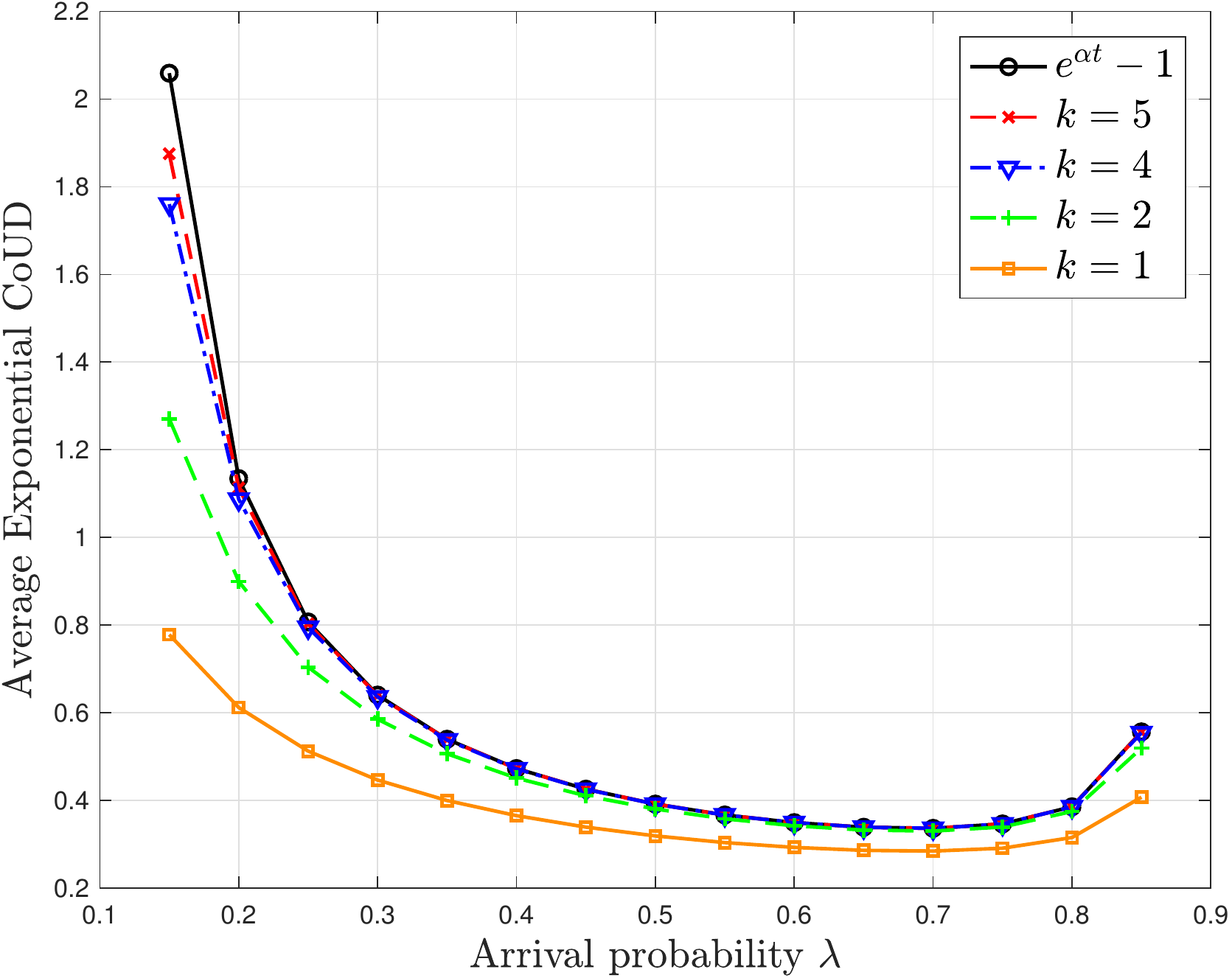}
	\caption{The average CoUD for the exponential function vs. the arrival probability $\lambda$, for the FCFS Geo/Geo/1 queue.}
	\label{fig:expCoUD_vs_lambda_geo_geo_1_Taylor}
\end{figure}

Fig.~\ref{fig:expCoUD_vs_lambda_geo_geo_1_Taylor} depicts the average CoUD for the exponential function in \eqref{eq:exp_Taylor_expan}, for the FCFS Geo/Geo/1 queue, with $\alpha=0.1$ and $\mu=0.9$.
We plot the exact result that is obtained using $f_s(t)=e^{\alpha t}-1$ together with the approximations that are obtained using a $k$th order polynomial.
We see that as the number of terms $k$ increases the approximation approached the result of the exact function. 
For $k=4$ the approximation is sufficiently good.
Using $t=15$, the parameter $GAP$ defined in Algorithm 1 ranges from $3.5$ for $k=1$ to $0.02$ for $k=5$.

\vspace*{-0.5cm}
\section{Final Remarks}
\label{sec:conclusions}

In this work, we have investigated a sample path of the AoI stochastic process and we have provided a general framework that establishes a relation among 
the AoI, the system delay, and the PAoI.
Our first aim is a complete characterization of the AoI process by obtaining its stationary distribution.
Our second aim is to be able to analyze any non-linear function of AoI and provide a wide range of potential uses of information ageing depending on the application.
Assuming ergodicity, we apply our results to three different discrete time queueing systems and we obtain general formulas of the stationary distributions and the z-transforms of the AoI and PAoI metrics.
These results can be utilized to provide AoI and PAoI performance guarantees, in terms of violation probabilities.
Next, we illustrate how our results can be used to obtain closed-form expressions of non-linear functions of the average AoI by providing some examples. 
In that direction, we use representations of functions as power series and develop an algorithm that approached the exact time-average performance up to a desired accuracy.
Finally, we consider the problem of maximizing freshness by obtaining the optimal arrival probability at the system, and discuss the universality of the solution with respect to the different cost functions.
The relation between AoI and PAoI, and the benefits of the preemptive LCFS queue discipline are further highlighted by the numerical results. 

\appendices

\section{Proof of Theorem \ref{theorem1}}\label{Appendix_A'}
The summation in \eqref{eq:age_fre_distribution} can be rewritten as a sum of disjoint parts. 
Starting from $t = 0$, the summation is decomposed into $Q_n(x)$ for $n = 1, \cdots, N(\mathcal{T})$, defined as
\begin{equation}
	Q_n(x) = \sum_{t=t'_n}^{t'_{n+1}} \mathbbm{1}_{\{ \Delta_t\leq x\}},
	\label{eq:Q_i_def}
\end{equation}	
and the areas of width $t'_1$ and $\mathcal{T}-t'_{N(\mathcal{T})}$ that we denote $\tilde{Q}$, defined as
\begin{equation}
\tilde{Q}(x) = \sum_{t=0}^{t'_{1}} \mathbbm{1}_{\{ \Delta_t\leq x\}}+ \sum_{t=t'_{N(\mathcal{T})}}^{\mathcal{T}} \mathbbm{1}_{\{ \Delta_t\leq x\}}.
\label{eq:Q_tilde}
\end{equation}
Using the definition of $\Delta_t$ in \eqref{eq:age_def} we have 
\begin{align}
Q_n(x) &= \sum_{t=t'_n}^{t'_{n+1}} \mathbbm{1}_{\{ \Delta_t\leq x\}} = \sum_{t=t'_n}^{t'_{n+1}} \mathbbm{1}_{\{  T_{n-1} + (t-t'_{n-1}) \leq x\}} = \sum_{u=T_n}^{A_{n+1}} \mathbbm{1}_{\{  u \leq x\}} \notag \\ 
%&= \sum_{t=t'_n}^{t'_{n+1}} \mathbbm{1}_{\{  T_{n-1} + (t-t'_{n-1}) \leq x\}}  = \notag \\
& = \sum_{u=0}^{A_{n+1}} \mathbbm{1}_{\{  u \leq x\}} - \sum_{u=0}^{T_{n}} \mathbbm{1}_{\{  u \leq x\}} = \sum_{u=0}^{x} \mathbbm{1}_{\{  T_n \leq u\}} - \sum_{u=0}^{x} \mathbbm{1}_{\{  A_{n+1} \leq u\}},
\label{eq:Q_n}
\end{align}
where the last equality follows from the following relation. For $r\geq 0$ and $ x\geq 0$,
\begin{equation}
	\sum_{u=0}^{r} \mathbbm{1}_{\{ u\leq x\}} = \min(r,x)
	=\sum_{u=0}^{x} (1-\mathbbm{1}_{\{ r\leq u\}}). 
	\label{eq:relation}
\end{equation}	

Then, the decomposition yields
\begin{align}
\Delta^{\dagger}(x) &= \lim_{\mathcal{T} \to \infty}   \frac{1}{\mathcal{T}} \sum_{t=0}^{\mathcal{T}} \mathbbm{1}_{\{ \Delta_t\leq x\}} = \lim_{\mathcal{T} \to \infty}   \frac{1}{\mathcal{T}} \left( \sum_{n=1}^{N(\mathcal{T})} Q_n(x) + \tilde{Q}(x) \right) = \notag \\
&=\lim_{\mathcal{T} \to \infty} \left(   \frac{N(\mathcal{T})}{\mathcal{T}} \frac{1}{N(\mathcal{T})}  \sum_{n=1}^{N(\mathcal{T})} Q_n(x)  + \frac{\tilde{Q}(x) }{\mathcal{T} } \right) = \lambda^{\dagger}  \sum_{u=0}^{x}  (T^{\dagger}(u) - A^{\dagger}(u)).
\label{eq:main_res}
\end{align}
Note that the term $\tilde{Q}(x)/\mathcal{T}$ goes to zero as $\mathcal{T}$ grows.
In addition, 
%assuming $\lim_{n \to \infty} n^{-1} t'_n =  1/\lambda^{\dagger}$,
from Lemma 1 we have 
\begin{equation}
\lim_{\mathcal{T} \to \infty} \frac{N(\mathcal{T})}{\mathcal{T}} =  \lambda^{\dagger}.
\label{eq:lambda}
\end{equation}

\section{Proof of Theorem \ref{theorem2}}\label{Appendix_B'}
Under the assumption of stationarity and ergodicity, Lemma 2 together with (5), and (6) yield the stationary distributions 
\begin{equation}
A(x)= \lim_{N \to \infty}  \frac{1}{N}\sum_{n=1}^{N} \mathbbm{1}_{\{ A_n\leq x\}}, \quad x \geq 0,
%\label{eq:peakage_fre_distribution}
\end{equation}

\begin{equation}
T(x)= \lim_{N \to \infty} \frac{1}{N}\sum_{n=1}^{N} \mathbbm{1}_{\{ T_n\leq x\}}, \quad x \geq 0.
%\label{eq:T_fre_distribution}
\end{equation}
Note that in this context of discrete time queueing systems $x$ cannot be less that one (time-slot).

We define the shifts $\mathscr{A}(x+1) =A(x)$ and $\mathscr{T}(x+1)= T(x)$, and the corresponding z-transforms

\begin{equation}
	\mathscr{A}^*(z)= \sum_{n'=1}^{\infty} \mathscr{A}(n') z^{n'}, \quad \text{and} \quad \mathscr{T}^*(z)= \sum_{{n'}=1}^{\infty} \mathscr{T}(n') z^{n'}.
\end{equation}

Moreover, let $n=n'-1$.
It then follows from Theorem 1 and the definition of z-transform in (10) that 

\begin{align}
	\Delta^{\ast}(z) &= \lambda \sum_{n=0}^{\infty} \sum_{u=0}^{n}  (\mathscr{T}(u+1) - \mathscr{A}(u+1)) z^{n+1} = \lambda \sum_{n=0}^{\infty} \sum_{u=0}^{n}  (T(u) - A(u)) z^{n} z \notag \\
	%&= \lambda \sum_{n=0}^{\infty} \sum_{u=0}^{n}  (T(u) - A(u)) z^{n+1} = \lambda z \sum_{n=0}^{\infty} \sum_{u=0}^{n}  (T(u) - A(u)) z^{n}  \notag \\
	&= \lambda z\left( \sum_{n=0}^\infty (T(n)-A(n)) z^n + z\sum_{n=0}^\infty (T(n)-A(n)) z^n + z^2\sum_{n=0}^\infty (T(n)-A(n)) z^n +...\right) \notag \\
	&= \lambda \left( T^*(z)-A^*(z) \right) z (1+z+z^2+...) \notag \\
	&=\lambda \left( T^*(z)-A^*(z) \right) \frac{z}{(1-z)}.
	%\label{eq:AoI_z_transform}
\end{align}
Thus, we have the z-transform of AoI in (11).

\section{Proof of Theorem \ref{theorem3}}\label{Appendix_C'}
Let $\bar{\rho}=\lambda(1-\mu)/\mu(1-\lambda)$.
Since $Y$ is geometrically distributed with parameter $\lambda$ we know that 
%$Y(x) =\text{Pr} (Y \leq x)=1-(1-\lambda)^x$.
\begin{equation}
Y(x) =\text{Pr} (Y \leq x)=1-(1-\lambda)^x.
\label{eq:Y_pdf}
\end{equation}
To derive the pmf and PDF of the system time $T$, we use the fact that the sum of $N$ geometric random variables where $N$ is geometrically distributed is also geometrically distributed, according to the convolution property of their generating functions \cite{Nelson2013probability}.
Let $S_j$, $j = 1,2, .. $ be independent and identically distributed geometric random variables with parameter $\mu$.
If an arriving packet sees $N$ packets in the system, then, the system time of that packet, using the memoryless property, can be written as the random sum $T = S_1 + \dots + S_{N}$.
To calculate the probability generating function of $T$ we condition on $N=n$ which occurs with probability $(1-\bar{\rho})\bar{\rho}^{n-1}$ and obtain 
\begin{align}
	T^{\ast}(z) &= \sum_{n=1}^{\infty} \left( \frac{\mu z}{1-(1-\mu)z}\right) ^n (1-\bar{\rho}) \bar{\rho}^{n-1} = \frac{\mu (1-\bar{\rho}) z}{1-(1-\mu (1-\bar{\rho}))z}.
	\label{eq:G_T}
\end{align}
This implies that the system time pmf is given by 
\begin{equation}
	P_T(x) = \mu (1-\bar{\rho}) (1-\mu+\mu \bar{\rho})^{x-1}.
	\label{eq:f_T}
\end{equation}
Therefore, the PDF of the system time is 
%$T(x) =\text{Pr} (T \leq x)=1-(1-\mu(1-\rho))^x$.
\begin{equation}
T(x) =\text{Pr} (T \leq x)=1-(1-\mu(1-\bar{\rho}))^x.
\label{eq:T_pdf}
\end{equation}
Finally, we know that 
\begin{equation}
	S^{\ast}(z) = \frac{\mu z}{1-(1-\mu)z}.
	\label{eq:S_z_stansform}
\end{equation}
Substituting all the relevant expressions to \eqref{eq:PAoI_z_transform} we obtain the result in  \eqref{eq:PAoI_gen_func}.

To obtain the z-transform of the AoI we utilize Theorem 2, together with \eqref{eq:PAoI_gen_func} and 
\eqref{eq:G_T}.

\section{Proof of Theorem \ref{theoremPreempt}}\label{Appendix_D'}
Since $Y$ is geometrically distributed with parameter $\lambda$ we 
%know that $Y(x) =\text{Pr} (Y \leq x)=1-(1-\lambda)^x$.
utilize \eqref{eq:Y_pdf}.
Similarly, for the service time $S$ with parameter $\mu$ we have 
%$S(x) =\text{Pr} (S \leq x)=1-(1-\mu)^x$.
\begin{equation}
S(x) =\text{Pr} (S \leq x)=1-(1-\mu)^x.
\label{eq:S_pdf}
\end{equation}
We apply \eqref{eq:zeta_def}, \eqref{eq:SlessthanY_def}, \eqref{eq:YlessthanS_def}, and \eqref{eq:YmorethanS_def}, accordingly to obtain
\begin{equation}
\theta = \sum_{n=1}^{\infty} (1-(1-\lambda)^n) P_S(n) =  S^{\ast}(1) - S^{\ast}(1-\lambda),
\label{eq:zeta_def_Geo}
\end{equation}	

\begin{equation}
S^{\ast}_{<Y}(z) = \frac{1}{1-\theta} \sum_{n=1}^{\infty} (1-\lambda)^n P_S(n) z^n =\frac{1}{1-\theta}  S^{\ast}( (1-\lambda) z),
\label{eq:SlessthanY}
\end{equation}	

\begin{equation}
Y^{\ast}_{<S}(z) = \frac{1}{\theta} \sum_{n=1}^{\infty} (1-\mu)^n P_Y(n) z^n = \frac{1}{\theta}  Y^{\ast}( (1-\mu) z), 
\label{eq:YlessthanS}
\end{equation}	

\begin{equation}
Y^{\ast}_{>S}(z) = \frac{1}{1-\theta} \sum_{n=1}^{\infty} (1-(1-\mu)^n) P_Y(n) z^n = \frac{1}{1-\theta}  \big[ Y^{\ast}(z) - Y^{\ast}( (1-\mu) z) \big]. 
\label{eq:YmorethanS}
\end{equation}	

The z-transform property that has been utilized is
\begin{equation}
\sum_{n=0}^{\infty} a^n \gamma(n) z^n = \Gamma^*(a z).
\label{eq:z_property}
\end{equation}	

Substituting all the relevant expressions to \eqref{eq:PAoI_z_preemption} and \eqref{eq:th2_preemption} we obtain the results in \eqref{eq:PAoI_gen_func_Pree} and \eqref{eq:AoI_gen_func_Pree}, respectively.

%\section*{Acknowledgment}
%The research leading to these results has been partially funded by the
%European Union's Horizon 2020 research and innovation programme under
%the Marie Sklodowska-Curie Grant Agreement No. 642743 (WiVi-2020).

%\newpage

\bibliography{references}
\bibliographystyle{IEEEtran}

\end{document}

%% file: age_vs_time_slotted.tex
\begin{tikzpicture}[scale=1.00]
% horizontal axis
\draw[->] (0,0) -- (8.2,0) node[anchor=north] {$t$};
% vertical axis
\draw[->] (0,0) -- (0,3.5) node[anchor=east] {$\Delta_t$};

%\draw	(-0.5,2) node[rotate=90] {Age};
\draw	(-0.3,0.38) node[anchor=south]{{\footnotesize $T_0$}};

%shadow area
 
%  \draw[fill=gray!10]  (1.5,0) -- (1.5,0.5) -- (2.0,0.5) -- (2.0,1)-- (2.5,1) -- (2.5,1.5)-- (3.0,1.5)-- (3.0,2)-- (3.5,2.0)-- (3.5,2.5)-- (4.0,2.5)-- (4.0,3.0)-- (4.5,3.0)-- (4.5,2.5)-- (4.0,2.5)-- (4.0,2.0)-- (3.5,2)-- (3.5,1.5)-- (3.0,1.5)-- (3.0,1) -- (2.5,1)-- (2.5,0.5)-- (2,0.5)-- (2,0);

%J1
\draw[fill=gray!20] (0,0.0) -- (0,1.0) -- (0.5,1)-- (0.5,1.5)-- (0.5,1.5)-- (1.0,1.5)-- (1.0,0.5) -- (0.5,0.5) -- (0.5,0.0);
 
 %J3  pattern=north west lines, pattern color=gray!50
\draw[fill=gray!20]  (1.5,0) -- (1.5,0.5) -- (2.0,0.5) -- (2.0,1)-- (2.5,1) -- (2.5,1.5)-- (3.0,1.5)-- (3.0,2)-- (3.5,2.0)-- (3.5,2.5)-- (4.0,2.5)-- (4.0,3.0)-- (4.5,3.0)-- (4.5,2.5)-- (4.0,2.5)-- (4.0,2.0)-- (3.5,2)-- (3.5,1.5)-- (3.0,1.5)-- (3.0,1) -- (2.5,1)-- (2.5,0.5)-- (2,0.5)-- (2,0);

%J2
\draw[pattern=crosshatch dots, pattern color=gray!20] (0.5,0)-- (0.5,0.5)-- (1,0.5)--   (1.0,1)-- (1.5,1) -- (1.5,1.5)-- (2.0,1.5)-- (2.0,2)-- (2.5,2.0)-- (2.5,1.0)-- (2.0,1.0)-- (2.0,0.5)-- (1.5,0.5)-- (1.5,0.0);

%J4
\draw[pattern=crosshatch dots, pattern color=gray!20]  (2.0,0.0) -- (2.0,0.5) -- (2.5,0.5) -- (2.5,1.0) -- (3.0,1.0) -- (3.0,1.5) -- (3.5,1.5) -- (3.5,2.0) -- (4.0,2.0) -- (4.0,2.5) -- (4.5,2.5) -- (4.5,3.0) -- (5.0,3.0) -- (5.0,3.5)  -- (5.5,3.5) -- (5.5,1.5) -- (5.0,1.5) -- (5.0,1.0) -- (4.5,1.0) -- (4.5,0.5) -- (4.0,0.5)  -- (4.0,0.0);

%White line
\draw[white] (5.5,3.5) -- (5.5,1.5) ;

%Jn
\draw[fill=gray!20] (6.0,0.0) -- (6.0,0.5) --  (6.5,0.5) -- (6.5,1)-- (7.0,1) -- (7.0,1.5)-- (7.5,1.5)-- (7.5,0.5) -- (7.0,0.5) -- (7.0,0.0);

% labels
\draw	(-0.5,0) node[anchor=north] {$t_0$}
           (0.5,0) node[anchor=north] {$t_1$}
		    (1.5,0) node[anchor=north] {$t_2$}
		    (2,0) node[anchor=north] {$t_3$}
		    (4.0,0) node[anchor=north] {$t_4$}
		    (6,0) node[anchor=north] {$t_{n-1}$}
		    (7.0,0) node[anchor=north] {$t_n$};
		    
\draw[->,>=stealth]    (1,0) -- (1,-0.4) node[anchor=south,below] {$t'_1$};
\draw[->,>=stealth]  (2.5,0) -- (2.5,-0.4) node[anchor=south,below] {$t'_2$};
\draw[->,>=stealth]  (4.5,0) -- (4.5,-0.4) node[anchor=south,below] {$t'_3$};
\draw[->,>=stealth]   (7.5,0) -- (7.5,-0.4) node[anchor=south,below] {$t'_n$};

\draw	(0.75,1.7) node{{\footnotesize $A_1$}}
(2.25,2.2) node{{\footnotesize $A_2$}}
(4.25,3.2) node{{\footnotesize $A_3$}}
(7.25,1.7) node{{\footnotesize $A_n$}};

% dots
\draw[fill] (0.0,0.5) circle [radius=0.040]
(1.0,0.5) circle [radius=0.040] 
(2.5,1.0) circle [radius=0.040]
(4.5,2.5) circle [radius=0.040]
(7.5,0.5) circle [radius=0.040];

%\draw[fill] (0.75,0.7) node{{\scriptsize $T_1$}}
%(2.25,1.2) node{{\scriptsize $T_2$}}
%(4.25,2.2) node{{\scriptsize $T_3$}}
%(7.25,0.7) node{{\scriptsize $T_n$}};

\draw[fill] %(1.25,0.5) node{{\scriptsize $T_1$}}
%(2.75,0.8) node{{\scriptsize $T_2$}}
(0.75,0.7) node{{\footnotesize $T_1$}}
(2.25,1.2) node{{\footnotesize $T_2$}}
(4.75,2.5) node{{\footnotesize $T_3$}}
(7.75,0.5) node{{\footnotesize $T_n$}};

%\draw	(0.55,0.75) node{{\scriptsize $J_1$}}
%		    (1.5,0.75) node{{\scriptsize $J_2$}};
%           %(3.1,2.5) node{{\scriptsize $Q_3$}};
%\draw   (3.5,0.75) node{{\scriptsize $J_4$}}
%           (7,0.75) node{{\scriptsize $J_n$}};
% \draw   (7.25,0.25) node{{\scriptsize $\tilde{J}$}};
           
%\draw[<-] (3.3,1.6) to [out=95,in=250] (3.3,2.5) node [above] {{\scriptsize $J_3$}};           

%''line segments'' below
           
%\draw [thick](0.5,-1.2) -- (1.5,-1.2) node[pos=.5,sloped,below] {$Y_2$} ;
%\draw[thick]  (0.5,-1.3) -- (0.5,-1.1); 
%\draw [thick](1.5,-1.2) -- (2.5,-1.2) node[pos=.5,sloped,below] {$T_2$} ;
%\draw[thick]  (1.5,-1.3) -- (1.5,-1.1) 
%                    (2.5,-1.3) -- (2.5,-1.1);
%                    
%\draw [thick](6,-1.2) -- (7.0,-1.2) node[pos=.5,sloped,below] {$Y_n$} ;
%\draw[thick]  (6,-1.3) -- (6,-1.1); 
%\draw [thick](7.0,-1.2) -- (7.5,-1.2) node[pos=.5,sloped,below] {$T_n$} ;
%\draw[thick]  (7.0,-1.3) -- (7.0,-1.1) 
%                    (7.5,-1.3) -- (7.5,-1.1);

%J1
%\draw[thick] (0,0.5) -- (0.5,0.5) -- (0.5,1)-- (1,1);
\draw[thick] (0,0.5) -- (0,1.0) -- (0.5,1)-- (0.5,1.5)-- (0.5,1.5)-- (1.0,1.5)-- (1.0,1.0);
%\draw[thick] (0,0.5) -- (0.25,0.5) -- (0.25,0.75)-- (0.5,0.75)-- (0.5,1.0)-- (0.75,1.0)-- (0.75,1.25)-- (1.0,1.25);

%sawtooth
%J2
\draw[thick] (1.0,1)-- (1.5,1) -- (1.5,1.5)-- (2.0,1.5)-- (2.0,2)-- (2.5,2.0)-- (2.5,1.5);
%J3
\draw[thick] (2.5,1) -- (2.5,1.5)-- (3.0,1.5)-- (3.0,2)-- (3.5,2.0)-- (3.5,2.5)-- (4.0,2.5)-- (4.0,3.0)-- (4.5,3.0);
%J4
\draw[thick]  (4.5,2.5)-- (4.5,3.0)-- (5.0,3.0)-- (5.0,3.5)-- (5.5,3.5);

%Jn
\draw[thick] (6.5,0.5) -- (6.5,1)-- (7.0,1) -- (7.0,1.5)-- (7.5,1.5)-- (7.5,0.5);

%vertical lines
\draw[dotted] (1,0) -- (1,1.5);
\draw[dotted] (2.5,0) -- (2.5,2.0); %t'2
\draw[dotted] (4.5,0) -- (4.5,3.0); %t'3
\draw[dotted] (7.5,0) -- (7.5,0.5); %t'n
%diagonal lines
\draw[dotted] (-0.5,0) -- (-0.5,0.5)-- (0,0.5); %t0
\draw[dotted] (0.5,0) -- (0.5,0.5) -- (1.0,0.5) -- (1.0,1); %t1
\draw[dotted] (1.5,0) -- (1.5,0.5) -- (2.0,0.5) -- (2.0,1)-- (2.5,1); %t2
\draw[dotted] (2,0) -- (2,0.5) -- (2.5,0.5) -- (2.5,1)-- (3.0,1) -- (3.0,1.5)-- (3.5,1.5)-- (3.5,2)-- (4.0,2.0)-- (4.0,2.5); %t3

\draw[dotted] (4,0) -- (4,0.5) -- (4.5,0.5)-- (4.5,1.0)-- (5.0,1.0); %t4
\draw[dotted] (6,0) -- (6,0.5) -- (6.5,0.5); %t'n-1
\draw[dotted] (7.0,0) -- (7.0,0.5) -- (7.5,0.5); %tn

%infinity symbol
\draw[thick]  (5.2,-0.15) -- (5.2,0.15) 
                    (5.3,-0.15) -- (5.3,0.15);
\draw[white, fill=white!50] (5.21,-0.2) -- (5.21,0.2) -- (5.29,0.2) -- (5.29,-0.2) ;                   

%%circles for A_n
%\draw[fill=white] (0.5,1.5) circle [radius=0.045] 
%(2.0,2.0) circle [radius=0.045]
%(4.0,3.0) circle [radius=0.045]
%(7.0,1.5) circle [radius=0.045];

%circles for A_n
\draw[fill=white] (1.0,1.5) circle [radius=0.045] 
(2.5,2.0) circle [radius=0.045]
(4.5,3.0) circle [radius=0.045]
(7.5,1.5) circle [radius=0.045];

\end{tikzpicture}

%% file: Draft.bbl
\begin{thebibliography}{10}
\providecommand{\url}[1]{#1}
\csname url@samestyle\endcsname
\providecommand{\newblock}{\relax}
\providecommand{\bibinfo}[2]{#2}
\providecommand{\BIBentrySTDinterwordspacing}{\spaceskip=0pt\relax}
\providecommand{\BIBentryALTinterwordstretchfactor}{4}
\providecommand{\BIBentryALTinterwordspacing}{\spaceskip=\fontdimen2\font plus
\BIBentryALTinterwordstretchfactor\fontdimen3\font minus
  \fontdimen4\font\relax}
\providecommand{\BIBforeignlanguage}[2]{{%
\expandafter\ifx\csname l@#1\endcsname\relax
\typeout{** WARNING: IEEEtran.bst: No hyphenation pattern has been}%
\typeout{** loaded for the language `#1'. Using the pattern for}%
\typeout{** the default language instead.}%
\else
\language=\csname l@#1\endcsname
\fi
#2}}
\providecommand{\BIBdecl}{\relax}
\BIBdecl

\bibitem{Kosta20_ICC}
A.~Kosta, N.~Pappas, A.~Ephremides, and V.~Angelakis, ``Non-linear age of
  information in a discrete time queue: Stationary distribution and average
  performance analysis,'' in \emph{Proc. IEEE ICC}, June 2020, pp. 1--6.

\bibitem{NET-060}
A.~Kosta, N.~Pappas, and V.~Angelakis, ``Age of information: A new concept,
  metric, and tool,'' \emph{Foundations and
  Trends$\textsuperscript{\textregistered}$ in Networking}, vol.~12, no.~3, pp.
  162--259, 2017.

\bibitem{Abd19_Magazine}
M.~A. Abd-Elmagid, N.~Pappas, and H.~S. Dhillon, ``On the role of age of
  information in the internet of things,'' \emph{IEEE Communications Magazine},
  vol.~57, no.~12, pp. 72--77, 2019.

\bibitem{Sun19_Book}
Y.~Sun, I.~Kadota, R.~Talak, and E.~Modiano, ``Age of information: A new metric
  for information freshness,'' \emph{Synthesis Lectures on Communication
  Networks}, vol.~12, no.~2, pp. 1--224, 2019.

\bibitem{Kaul12_INFOCOM}
S.~Kaul, R.~Yates, and M.~Gruteser, ``Real-time status: How often should one
  update?'' in \emph{Proc. IEEE INFOCOM}, March 2012, pp. 2731--2735.

\bibitem{Kaul12_CISS}
S.~K. Kaul, R.~D. Yates, and M.~Gruteser, ``Status updates through queues,'' in
  \emph{Proc. IEEE CISS}, March 2012, pp. 1--6.

\bibitem{Najm16_ISIT}
E.~Najm and R.~Nasser, ``Age of information: The gamma awakening,'' in
  \emph{Proc. IEEE ISIT}, July 2016, pp. 2574--2578.

\bibitem{Bedewy16_ISIT}
A.~M. Bedewy, Y.~Sun, and N.~B. Shroff, ``Optimizing data freshness,
  throughput, and delay in multi-server information-update systems,'' in
  \emph{Proc. IEEE ISIT}, July 2016, pp. 2569--2573.

\bibitem{Yates18_INFOCOM}
R.~D. Yates, ``Age of information in a network of preemptive servers,'' in
  \emph{Proc. IEEE INFOCOM Workshops}, April 2018, pp. 118--123.

\bibitem{Najm18_INFOCOM}
E.~Najm and E.~Telatar, ``Status updates in a multi-stream {M/G/1/1} preemptive
  queue,'' in \emph{Proc. IEEE INFOCOM}, April 2018, pp. 124--129.

\bibitem{Costa16}
M.~Costa, M.~Codreanu, and A.~Ephremides, ``On the age of information in status
  update systems with packet management,'' \emph{IEEE Transactions on
  Information Theory}, vol.~62, no.~4, pp. 1897--1910, April 2016.

\bibitem{Soysal19_arXiv}
A.~Soysal and S.~Ulukus, ``Age of information in {G/G/1/1} systems: Age
  expressions, bounds, special cases, and optimization,''
  \emph{arXiv:1905.13743}, 2019.

\bibitem{Kosta19_ISIT}
A.~Kosta, N.~Pappas, A.~Ephremides, and V.~Angelakis, ``Queue management for
  age sensitive status updates,'' in \emph{Proc. IEEE ISIT}, July 2019, pp.
  330--334.

\bibitem{Kosta19_JCN}
A.~Kosta, N.~Pappas, A.~Ephremides, and V.~Angelakis, ``Age of information
  performance of multiaccess strategies with packet management,''
  \emph{IEEE/KICS Journal of Communications and Networks}, vol.~21, no.~3, pp.
  244--255, 2019.

\bibitem{Kam16}
C.~Kam, S.~Kompella, G.~D. Nguyen, and A.~Ephremides, ``Effect of message
  transmission path diversity on status age,'' \emph{IEEE Transactions on
  Information Theory}, vol.~62, no.~3, pp. 1360--1374, March 2016.

\bibitem{Bedewy19_transactionsIT}
A.~M. Bedewy, Y.~Sun, and N.~B. Shroff, ``Minimizing the age of information
  through queues,'' \emph{IEEE Transactions on Information Theory}, vol.~65,
  no.~8, pp. 5215 -- 5232, 2019.

\bibitem{Tripathi19_arXiv}
V.~Tripathi, R.~Talak, and E.~Modiano, ``Age of information for discrete time
  queues,'' \emph{arXiv:1901.10463}, 2019.

\bibitem{Talak18_Determinacy}
R.~Talak, S.~Karaman, and E.~Modiano, ``Can determinacy minimize age of
  information?'' \emph{arXiv:1810.04371}, 2018.

\bibitem{Chen16_ISIT}
K.~Chen and L.~Huang, ``Age-of-information in the presence of error,'' in
  \emph{Proc. IEEE ISIT}, July 2016, pp. 2579--2583.

\bibitem{Yates18_arXivSHS}
R.~D. {Yates}, ``The age of information in networks: Moments, distributions,
  and sampling,'' \emph{IEEE Transactions on Information Theory}, vol.~66,
  no.~9, pp. 5712--5728, 2020.

\bibitem{Inoue19_transactionsIT}
Y.~Inoue, H.~Masuyama, T.~Takine, and T.~Tanaka, ``A general formula for the
  stationary distribution of the age of information and its application to
  single-server queues,'' \emph{IEEE Transactions on Information Theory},
  vol.~65, no.~12, pp. 8305 -- 8324, 2019.

\bibitem{Champati19_INFOCOM}
J.~P. Champati, H.~Al-Zubaidy, and J.~Gross, ``On the distribution of aoi for
  the {GI/GI/1/1} and {GI/GI/1/2*} systems: Exact expressions and bounds,'' in
  \emph{Proc. IEEE INFOCOM}, Apr. 2019, pp. 37--45.

\bibitem{Kesidis19_arXiv}
G.~Kesidis, T.~Konstantopoulos, and M.~Zazanis, ``The distribution of
  age-of-information performance measures for message processing systems,''
  \emph{Queueing Systems 95, 203-250}, 2020.

\bibitem{Devassy19_JSAC}
R.~Devassy, G.~Durisi, G.~C. Ferrante, O.~Simeone, and E.~Uysal, ``Reliable
  transmission of short packets through queues and noisy channels under latency
  and peak-age violation guarantees,'' \emph{IEEE Journal on Selected Areas in
  Communications}, vol.~37, no.~4, pp. 721--734, 2019.

\bibitem{Kosta17_ISIT}
A.~Kosta, N.~Pappas, A.~Ephremides, and V.~Angelakis, ``Age and value of
  information: Non-linear age case,'' in \emph{Proc. IEEE ISIT}, June 2017, pp.
  326--330.

\bibitem{Kosta19_arXiv}
A.~Kosta, N.~Pappas, A.~Ephremides, and V.~Angelakis, ``The cost of delay in
  status updates and their value: Non-linear ageing,'' \emph{IEEE Transactions
  on Communications}, vol.~68, no.~8, pp. 4905--4918, 2020.

\bibitem{Sun2017_transactions}
Y.~Sun, E.~Uysal-Biyikoglu, R.~D. Yates, C.~E. Koksal, and N.~B. Shroff,
  ``Update or wait: How to keep your data fresh,'' \emph{IEEE Transactions on
  Information Theory}, vol.~63, no.~11, pp. 7492--7508, 2017.

\bibitem{Sun18_JCN}
Y.~Sun and B.~Cyr, ``Sampling for data freshness optimization: Non-linear age
  functions,'' \emph{IEEE/KICS Journal of Communications and Networks},
  vol.~21, no.~3, pp. 204--219, 2019.

\bibitem{Sun18_SPAWC}
Y.~Sun and B.~Cyr, ``Information aging through queues: A mutual information
  perspective,'' in \emph{Proc. IEEE SPAWC}, June 2018, pp. 1--5.

\bibitem{Sun19_transactions}
Y.~Sun, Y.~Polyanskiy, and E.~Uysal, ``Sampling of the wiener process for
  remote estimation over a channel with random delay,'' \emph{IEEE Transactions
  on Information Theory}, vol.~66, no.~2, pp. 1118--1135, 2020.

\bibitem{Ornee19_WiOpt}
T.~Z. {Ornee} and Y.~{Sun}, ``Sampling for remote estimation through queues:
  Age of information and beyond,'' in \emph{Proc. WiOPT}, June 2019, pp. 1--8.

\bibitem{ElTaha1999sample}
M.~El-Taha and S.~Stidham~Jr, \emph{Sample-path analysis of queueing
  systems}.\hskip 1em plus 0.5em minus 0.4em\relax Springer Science \& Business
  Media, 1999.

\bibitem{Kleinrock}
L.~Kleinrock, \emph{{Queueing Systems}}.\hskip 1em plus 0.5em minus 0.4em\relax
  Wiley Interscience, 1975, vol. I: Theory.

\bibitem{Nelson2013probability}
R.~Nelson, \emph{Probability, stochastic processes and queueing theory: the
  mathematics of computer performance modeling}.\hskip 1em plus 0.5em minus
  0.4em\relax New York: Springer-Verlang, 1995.

\end{thebibliography}
